\documentclass[aps,prd,showpacs,notitlepage,nofootinbib,superscriptaddress,floatfix,showkeys,twocolumn]{revtex4-1}
\usepackage{mathrsfs} 
\usepackage[utf8]{inputenc}
\usepackage[T1]{fontenc}
\usepackage{upgreek}
\usepackage{float}
\usepackage{comment}
\usepackage{cancel}
\usepackage[normalem]{ulem} 
\usepackage{soul}
\usepackage{siunitx}
\usepackage{graphicx}
\usepackage{gensymb}
\usepackage{amsmath, cases}
\usepackage[usenames]{color}
\usepackage{amssymb}
\usepackage{hyperref}
\usepackage{wrapfig}
\usepackage{cleveref}
\newcommand{\be}{\begin{equation}}
\newcommand{\ee}{\end{equation}}
\newcommand{\bea}{\begin{aligned}}
\newcommand{\eea}{\end{aligned}}
\newcommand{\pr}{\partial}

\newcommand{\bse}{\begin{subequations}}
\newcommand{\ese}{\end{subequations}}

\newcommand{\rb}{\bf r}
\newcommand{\xb}{\bf x}

\newcommand{\bmm}{\begin{multline}}
\newcommand{\emm}{\end{multline}}

\newcommand{\mi}{\mathrm{i}}
\numberwithin{equation}{section}
\renewcommand{\thesection}{\Roman{section}}
\renewcommand{\theequation}{\arabic{section}.\arabic{equation}}
\begin{document}
\title{Absorption cross section of a Schwarzschild black hole for a massive vector field}
\author{Kaustubh Mukund Vispute}
\email{kaustubhvispute28@gmail.com}
\affiliation{Department of Physics, 
Indian Institute of Technology Guwahati, Assam 781039, India}
\author{Rajesh Karmakar}
\email{rajesh@shu.edu.cn}
\affiliation{Department of Physics, Shanghai University, 99 Shangda Road, Shanghai, 200444, China}

\begin{abstract}
In the present work, we study the absorption cross section of a Schwarzschild black hole for a massive vector field over arbitrary frequencies. Working in the Frolov--Krtou\v{s}--Kubiz\v{n}\'ak--Santos (FKKS) basis, we show how the conserved flux of the normalized Proca field naturally leads to the usual definitions of the absorption cross section in terms of the reflection and transmission coefficients. We then numerically compute these quantities over arbitrary frequencies. In contrast to the massless (photon) case, massive vector bosons exhibit new features arising from the field mass. In particular, the mass term introduces a longitudinal degree of freedom in addition to the transverse modes. Furthermore, it leads to a scalar-type branch in the even parity transverse modes, thereby breaking the usual degeneracy with the odd parity sector found in the massless case. We illustrate how these characteristics manifest themselves in the transmission and absorption spectrum. Given the recent developments in the study of ultralight bosonic fields in black hole spacetimes, the present analysis of the transmission properties of massive vector bosons bears particular significance.
\end{abstract}

\maketitle

\newpage
\section{Introduction}\label{intro}
Massive vector fields, commonly referred to as Proca, arise in multiple theoretical frameworks, including modified gravity theories with vector–tensor interactions \cite{DeFelice:2016yws}, extensions of the Standard Model \cite{Langacker:2008yv, Essig:2013lka}, and various contexts in string theory compactifications \cite{Anchordoqui:2020tlp, Goodsell:2010ie}. Specifically, the ultralight ones are considered promising candidates for dark matter \cite{Nelson:2011sf, Arvanitaki:2010sy}. Being massive, they propagate with an additional longitudinal degree of freedom, leading to interesting gravitational interactions as compared to the massless case. In strong-field regimes, such as near the black holes (BHs), which are now confirmed to be situated at the centers of galaxies through gravitational wave detections \cite{LIGOScientific:2016aoc} and BH shadow image observations \cite{EventHorizonTelescope:2019dse, EventHorizonTelescope:2022wkp}, are also believed to be embedded in a DM medium, whose density profile described by a broken power law with an inner central spike \cite{Battaner:2000ef, Lacroix:2018zmg, Shen:2023kkm}. The nature of such dark matter is not known yet and could very well consist of massive vector bosons \cite{Karmakar:2025drp, Hancock:2025ois, Kitajima:2023fun, Zi:2024lmt}. As far as the stability and survival of such a distribution are concerned \cite{Traykova:2023qyv, Hancock:2025ois}, the scattering of the Proca field bears significance. On the other hand, recent studies have shown that the ringdown waveform, viewed as a superposition of quasi-normal modes, encodes the greybody factor \cite{Oshita:2023cjz, Konoplya:2024lir}, thereby making the investigation of the scattering properties of BH spacetime of practical importance \cite{Rosato:2024arw, Oshita:2024wgt, Lo:2025njp}.  

Generally, the set of boundary conditions required to study the scattering, such as reflection and transmission fluxes, is different from that required for quasi-normal modes \cite{Chandrasekhar:1975zza}. However, in both cases, the separability of the field equation in the background spacetime of interest is of immense help. Significant progress in this direction has been achieved over the years, pioneered by Regge-Wheeler \cite{Regge:1957td}, Zerilli \cite{Zerilli:1970se}, and Press-Teukolsky \cite{Teukolsky:1972my, Teukolsky:1973ha, Press:1973zz}, among others \cite{Chandrasekhar:1975nkd}. Importantly, however, the standard Teukolsky framework works well primarily for massless fields \cite{Teukolsky:1972my}. While the extension to the massive case is relatively straightforward for scalar fields, the generalization to higher-spin fields is non-trivial. For a massive scalar field on Schwarzschild spacetime, it has been shown that the low-energy absorption cross section (ACS) deviates from the standard massless limit and exhibits a distinct mass-dependent behaviour \cite{Jung:2004yh}. Likewise, massive fermionic fields have been studied in the context of BH absorption and scattering \cite{Dolan:2006vj, Doran:2005vm}, showing that the inclusion of the mass can substantially modify the scattering properties. 

For a massive vector field, the mass term couples the components of the field in such a manner that the equations of motion do not decouple in a straightforward manner. Because of this complication, while the scattering of the electromagnetic (EM) field has been extensively studied in the Schwarzschild spacetime\cite{Crispino:2007qw, Crispino:2009xt}, the corresponding massive vector case has not been explored to the same extent. Partly owing to this complexity, the coupled differential equations have been solved numerically to study the transmission properties and the greybody factor associated with Hawking radiation\footnote{The Hawking flux is particularly relevant for smaller BH masses. Therefore, a precise estimate of the greybody factor, and thereby of the radiation flux, is also important in the context of primordial BHs with masses much smaller than the solar mass. In this context, there are useful numerical packages with BlackHAWK \cite{Arbey:2019mbc,Arbey:2021mbl, Chan:2022gqd} considering massless fields, whereas extension to the massive case not consistently done (see e.g., Frisbhee \cite{Cheek:2022dbx}).} \cite{Herdeiro:2011uu}. Concerning the stability of compact objects, the quasi-bound states and the quasinormal modes of the Proca field have been previously analyzed, and, compared to the massless case, are found to exhibit a non-degenerate spectrum between the even and odd parity sectors \cite{Galtsov:1984ixy, Konoplya:2005hr, Rosa:2011my, Dolan_2018}.  Through these analyses, the significance of the mass term in breaking the degeneracy of the two transverse degrees of freedom present in the usual Maxwell case, as well as in governing the dynamics of the longitudinal degrees of freedom, has been understood.  Recently, the FKKS method \cite{Frolov:2018ezx} has provided a remarkable strategy to circumvent the issues of separability of the equation of motion.

The focus of the present work is on the ACS of the Schwarzschild  BH for the Proca field. While we leave the extension to the rotating case for future work, the current analysis can be generalized straightforwardly to other static, spherically symmetric BHs with metric structures similar to that of the Schwarzschild \cite{Karmakar:2024hng}. In the present work, we consider the Proca in the test field approximation and implement the FKKS technique to decouple the wave equation. Along these lines, there has recently been an effort to obtain the reflection and transmission coefficients for Proca fields using the WKB approximation \cite{Bunjusuwan:2025enh}. We perform the whole analysis without resorting to such approximations. On the other hand, the ACS of a Schwarzschild BH for the Proca field has not yet been reported. Therefore, it is imperative to analyze the polarization-dependent ACS in this case. From this perspective, the present article may be viewed as an extension of earlier studies of massless vector fields in Schwarzschild BH spacetimes \cite{Crispino:2007qw}.


The rest of the paper is structured as follows. In \Cref{sec:proca_eqns}, the procedure to obtain the decoupled equation of motion for the Proca field has been outlined. The corresponding effective potentials for different parity modes have also been illustrated with plots in this section. Next, in \Cref{sec:numerical_recipe}, the numerical procedure to determine the reflection and transmission coefficients for different parity modes has been discussed. Then, we have provided the analysis for the normalization of the Proca field in \Cref{sec:normalization}. After that, in \Cref{sec:ACS_proca}, the total ACS of the Schwarzschild BH has been defined and analyzed for individual parity modes, as well as for their combined contribution. Finally, we have concluded with future outlook in \Cref{sec:conclusion}.

List of notations used: ${\rm m}, {\rm o}$ and ${\rm e(\pm)}$ stand for monopole, odd parity and even parity (positive and negative sectors) modes respectively.

Throughout the analysis, we will use the natural units: $\hbar=c=1$.
\section{Proca field in Schwarzschild spacetime}\label{sec:proca_eqns}
The Schwarzschild spacetime is described by the following line element in $x^\mu=\{t,r,\theta,\phi\}$,
\be
ds^2=-f(r)dt^2+f(r)dr^2+r^2d\theta^2+r^2\sin^2\theta d\phi^2,
\ee
where $f(r)=1-r_h/r$, with $r_h=2M$, $M$ representing the BH mass. Considering the Proca field to be a test field{\footnote{See Appendix D of \cite{Hancock:2025ois} for a discussion of the validity of this approximation in the context of dark matter described by a Proca field. In particular, it is argued that the approximation remains applicable for small and intermediate mass BHs, whereas for supermassive BHs, $M \geq 10^{10}M_{\odot}$, the backreaction cannot straightaway be neglected.}} and minimally coupled with the background spacetime, the equation of motion reads
\be\label{eq.eom1}
-\frac{1}{\sqrt{-g}}\pr_\mu\left(\sqrt{-g}F^{\mu\nu}\right)+\mu^2 A^\nu=0,
\ee
where $\mu$ stands for the Proca mass.
Given Ricci flat spacetime,  it follows from the field equation that 
\be\label{eq: gauge.cond}
\pr_\mu\left(\sqrt{-g}A^\mu\right)=0,
\ee
which is known as the Lorenz condition\footnote{See \cite{jackson1999classical} (page 294) for an interesting discussion on the nomenclature.}. Since this condition is automatically satisfied, it need not be imposed separately on the solutions. However, utilizing it can simplify the computation at various stages, as we will see in a later section. 

With a spin-$1$ field, it is convenient to implement the following set of vector spherical harmonics for describing the angular part of the Proca field \cite{Fernandes:2021qvr, Ge:2025yqk}:
\be
\bea
&Z^{(1)lm}_\mu=\left[1,0,0,0\right]Y^{lm}(\theta, \phi),\\
&Z^{(2)lm}_\mu=\left[0,f^{-1},0,0\right]Y^{lm}(\theta, \phi),\\
&Z^{(3)lm}_\mu=\frac{r}{\sqrt{L}}\left[0,0,\pr_\theta,\pr_\phi\right]Y^{lm}(\theta, \phi),\\
&Z^{(4)lm}_\mu=\frac{r}{\sqrt{L}}\left[0,0,\frac{1}{\sin\theta}\pr_\phi,-\sin\theta\pr_\theta\right]Y^{lm}(\theta, \phi),\\
\eea
\ee
where $L=l(l+1)$. Whereas $Y_{lm}(\theta, \phi)$ represents the usual scalar spherical harmonics. The above expressions satisfy the following orthogonality condition
\be\label{eq: orthogonality_cond}
\int\left(Z^{(i)lm}_\mu\right)^*\eta^{\mu\nu}\left(Z^{(i')l'm'}_\nu\right)\sin\theta d\theta d\phi=\delta_{ii'}\delta_{ll'}\delta_{mm'},
\ee
with 
\be
\eta_{\mu\nu}={\rm diag}[1,f^2,1/r^2,1/(r^2\sin^2\theta)].
\ee
Under parity transformation, $(\theta,\phi)\to (\pi-\theta, \phi+\pi)$, the above vector spherical harmonics transform as
\be
\bea
&Z^{(1,2,3)lm}=(-1)^lZ^{(1,2,3)lm},\\
&Z^{(4)lm}=(-1)^{(l+1)}Z^{(4)lm},
\eea
\ee
namely, even parity and odd parity respectively. On this basis, we decompose the vector field as
\be\label{eq: Proca_sph_decomp}
A_\mu(t,{\bf r})=\frac{1}{r}\sum^4_{i=1}\sum_{lm}c_{(i)}u^{lm}_{(i)}(t,r)Z^{(i)lm}_\mu(r, \theta, \phi),
\ee
with $c_{(1)}=c_{(2)}=1$ and $c_{(3)}=c_{(4)}=1/\sqrt{L}$. Notably, (${\bf r}$) stands for ($r,\theta,\phi$). Substituting the above decomposed field in the equation of motion \eqref{eq.eom1}, it results in a set of coupled equations in the even parity sector \cite{Bunjusuwan:2025enh},
\be\label{eq: even_coupled_eom}
\bea
&\hat{\mathcal{D}}_2u_{(1)}+\left[\frac{r_h}{r^2}\left(\pr_t{u}_{(2)}-\pr_{r_*}u_{(1)}\right)\right]=0,\\
&\hat{\mathcal{D}}_2u_{(2)}+\frac{1}{r^2}\left[r_h\left(\pr_t{u}_{(1)}-\pr_{r_*}u_{(2)}\right)-2f^2\left(u_{(2)}-u_{(3)}\right)\right]=0,\\
&\hat{\mathcal{D}}_2u_{(3)}+\left[\frac{2}{r^2}fLu_{(2)}\right]=0,\\
\eea
\ee
and 
a completely decoupled equation for the odd parity sector,
\be\label{eq: Odd_Parity_eom1}
\mathcal{D}_2u_{(4)}=0,
\ee
with 
\be
\hat{\mathcal{D}}_2\equiv -\pr^2_t+\pr^2_{r_*}-f\left[\frac{L}{r^2}+\mu^2\right].
\ee
Here and in the following discussion, we have used the Tortoise coordinate, defined by $dr_*=f^{-1}dr$. 

\begin{figure}[t!]
    \centering
\includegraphics[scale=0.55]{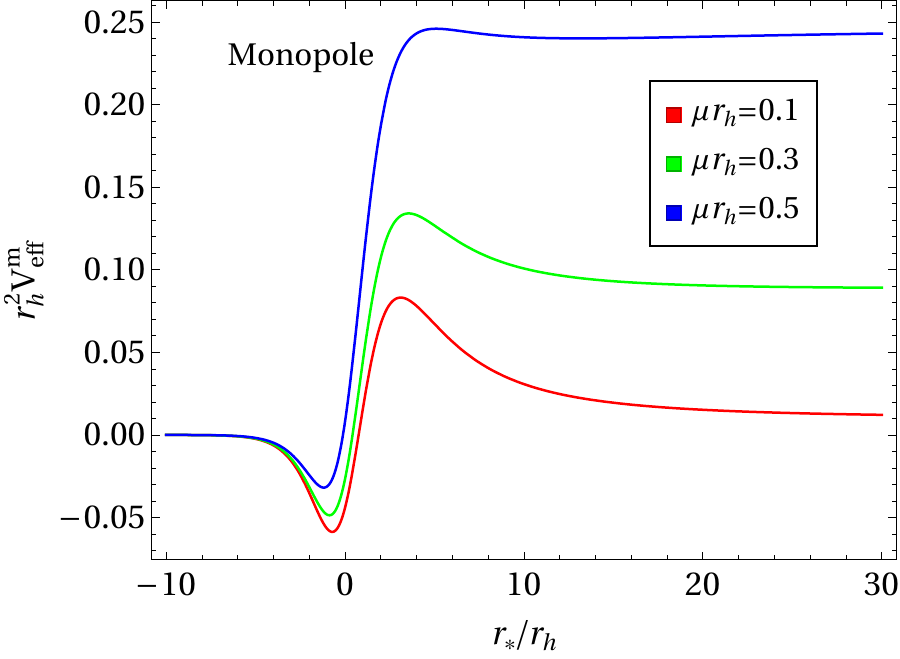}
\caption{Potential barrier for the monopole mode of the Proca field has been plotted with $r_*$. Variations have been shown for different values of field mass $\mu$, scaled with the BH horizon radius $r_h$.}\label{fig:monopole_pot}
\end{figure}

Let us begin with the $l=0$ mode, which is referred to as the monopole (${\rm m}$) mode and belongs to the even parity sector. The corresponding equation of motion reads
\be
-\pr^2_tu_{(2)}+\pr^2_{r_*}u_{(2)}-V^{\rm m}_{\rm eff}(r)u_{(2)}=0,
\ee
with the effective potential,
\be\label{eq: mono_pot}
V^{\rm m}_{\rm eff}(r)=f\left(\frac{2}{r^2}-\frac{3r_h}{r^3}+\mu^2\right).
\ee
Given the static nature of Schwarzschild spacetime, it is suitable to consider $u_{(i)}(t,r)=e^{-i\omega t}u_{(i)}(r, \omega)$. Therefore, the governing equation for the monopole mode, now, reads
\be
\frac{d^2u_{(2)}}{dr^2_*}+\left[\omega^2-V^{\rm m}_{\rm eff}(r)\right]u_{(2)}=0.
\label{eq. Monopole mode}
\ee
Interestingly, the effective potential for the monopole mode near the BH horizon happens to be negative and bounded from below, as evident from Fig.~\ref{fig:monopole_pot}. It is quite distinct compared to the monopole scalar field mode \cite{Hui:2019aqm}. In the latter case, the effective potential remains strictly positive throughout the exterior region of the BH \cite{sanchez1978absorption}. On the other hand, for the Proca monopole case, the effective potential contains a term proportional to $-3/r^3$ which dominates over the $2/r^2$ contribution at small radii (approximately for $r \lesssim 1.5\,r_h$). As a result, the effective potential develops a dip well just outside the event horizon. The implications of the monopole mode's unique nature in the quasinormal mode spectrum have been investigated previously \cite{Konoplya:2005hr}. Nevertheless, as the radial distance increases, the $2/r^2$ term and the mass term become dominant, causing the potential to increase and form a positive barrier before eventually decaying to a constant value at spatial infinity. Concerning the effect of the mass term, it is obvious from the expression of the effective potential \eqref{eq: mono_pot} that the potential barrier increases with the increment of the field mass, $\mu$, as has also been illustrated in Fig. \ref{fig:monopole_pot}. At this point, it is important to mention that considering the massless limit, $\mu\to 0$, to the monopole mode, makes it a pure gauge mode, which can be fixed to be zero by appropriate gauge choices (see e.g., \cite{Rosa:2011my}) as is expected in the Maxwell case.  

\begin{figure}[t!]
\includegraphics[scale=0.55]{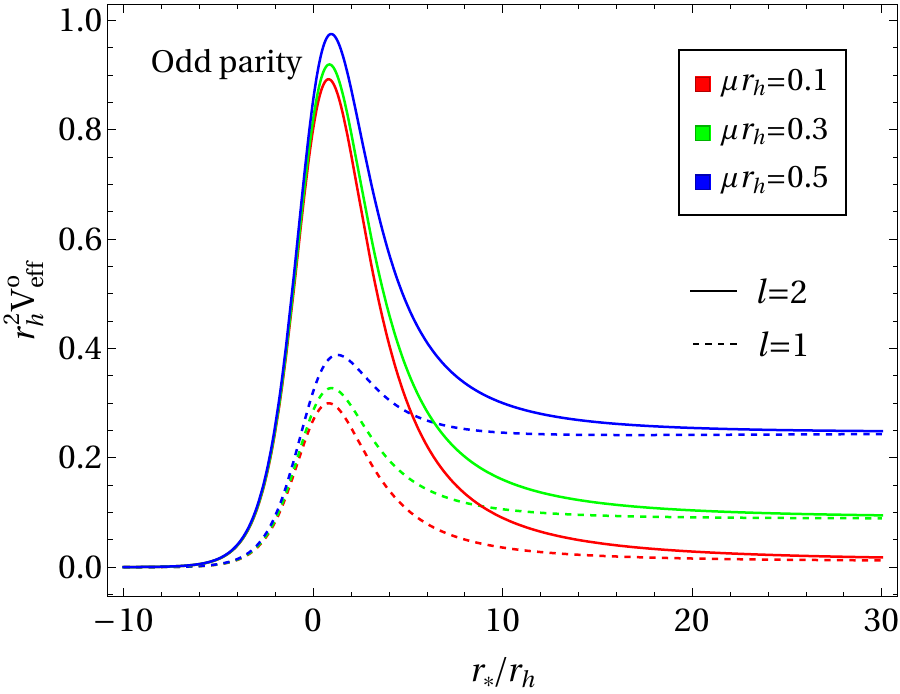}
\caption{Potential barrier for the odd parity mode of the Proca field has been plotted with $r_*$. Variations have been shown for different values of field mass $\mu$ by scaling with BH horizon radius $r_h$. Whereas the dotted and solid lines represent two consecutive orbital modes, $l=1$ and $l=2$, respectively.}\label{fig:odd_pot}
\end{figure}

Coming to the case of non-zero $l$ modes, separating the time part in the similar manner as discussed previously, we arrive at
\be\label{eq: Odd_Parity_eom2}
\frac{d^2u_{(4)}}{dr^2_*}+\left[\omega^2-V^{\rm o}_{\rm eff}(r)\right]u_{(4)}=0,
\ee
with the effective potential, for the odd parity (${\rm o}$), given by 
\be\label{eq:Odd_pot}
V^{\rm o}_{\rm eff}(r)=f\left(\frac{L}{r^2}+\mu^2\right).
\ee
It is obvious from the above expression that, unlike the monopole case, the effective potential for the odd parity mode is positive definite throughout the entire exterior region of the BH and vanishes at the horizon. As illustrated in the Fig.~\ref{fig:odd_pot}, the height of the potential barrier increases with increasing orbital mode, $l$. However, the increase in the peak of the potential barrier is comparatively smaller with increasing field mass, $\mu$. 

Concerning the even parity sector, with $l\neq 0$, the equations of motion \eqref{eq: even_coupled_eom} are not straightforwardly decoupled. Hence, we utilize the Frolov--Krtou\v{s}--Kubiz\v{n}\'ak--Santos (FKKS) method \cite{Frolov:2018ezx}, as has been previously implemented in \cite{Percival:2020skc, Fernandes:2021qvr}, which parametrizes the components as
\be\label{eq: fkks_ansatz}
\bea
&u_{(1)}(\omega, r)=-\frac{ifr(\nu r\pr_r +\omega/f)}{q_r}\bar{R}(\omega, r),\\
&u_{(2)}(\omega, r)=\frac{fr(\pr_r-\omega\nu r/f)}{q_r}\bar{R}(\omega, r),\\
&u_{(3)}(\omega, r)=L\bar{R}(\omega, r),
\eea
\ee
with $q_r=1+\nu^2r^2$. Where $\nu$ represents a separation constant and can be determined in the following steps. By inserting the expressions for $u_{(2)}$ and $u_{(3)}$ into the final equation of \eqref{eq: even_coupled_eom}, the differential equation governing $\bar{R}$ turns out as
\be
\frac{d^2\bar{R}}{dr^2_*}+\frac{2f}{rq_r}\frac{d\bar{R}}{dr_*}+\left[\omega^2-f\left(\frac{L}{r^2}+\mu^2+\frac{2\omega\nu}{q_r}\right)\right]\bar{R}=0.
\ee
On the other hand, utilizing all the components of \eqref{eq: fkks_ansatz} in the Lorenz condition \eqref{eq: gauge.cond}, we obtain another form of the above differential equation governing the same $\bar{R}$,
\be
\frac{d^2\bar{R}}{dr^2_*}+\frac{2f}{rq_r}\frac{d\bar{R}}{dr_*}+\left[\omega^2-f\left(3\omega\nu+\frac{q_rL}{r^2}-\frac{2\omega\nu^3r^2}{q_r}\right)\right]\bar{R}=0.
\ee
Therefore, the preceding two equations must match, which leads to the following conditions involving the separation constant,
\be\label{eq: lnumu_cond}
L\nu^2+\omega \nu-\mu^2=0.
\ee
The roots of the above equation reads
\be\label{eq: nuplusminus}
\nu_{\pm}=-\frac{\omega}{L}\left[\frac{1\pm \sqrt{1+4L\mu^2/\omega^2}}{2}\right].
\ee
\begin{figure*}[t]
\centering
\includegraphics[width=\textwidth]{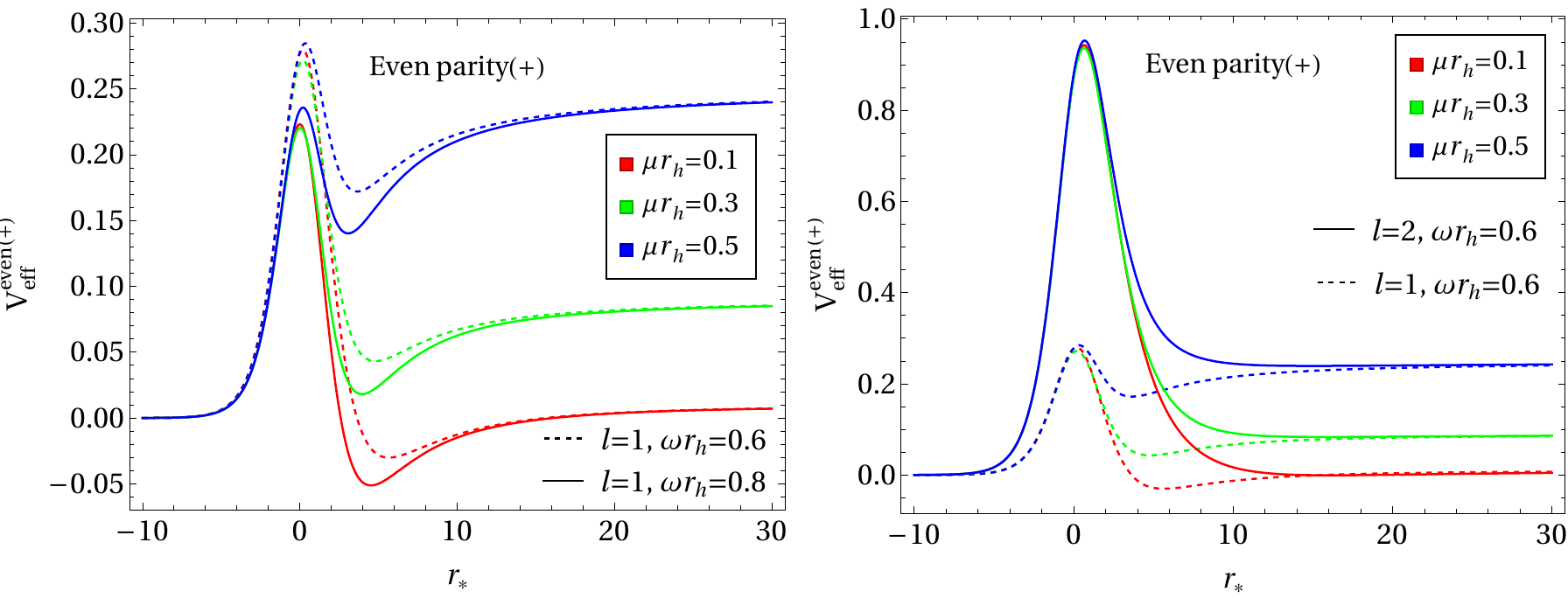}
\caption{The effective potential barrier for the positive even parity mode of the Proca field is plotted as a function of $r_*$. Since the effective potential depends on the frequency, the {\bf left panel} shows the potential for a fixed orbital number, $l=1$, and two different frequencies, $\omega r_h=0.6$ and $\omega r_h=0.8$, represented by dotted and solid lines, respectively. In the {\bf right panel}, the frequency is held fixed, while two consecutive orbital modes, $l=1$ and $l=2$, are shown, again represented by dotted and solid lines, respectively. In both panels, the variation with the field mass $\mu$ is illustrated using different colours.}\label{fig:evenplus_pot}
\end{figure*}
\begin{figure*}[t]
\centering
\includegraphics[width=\textwidth]{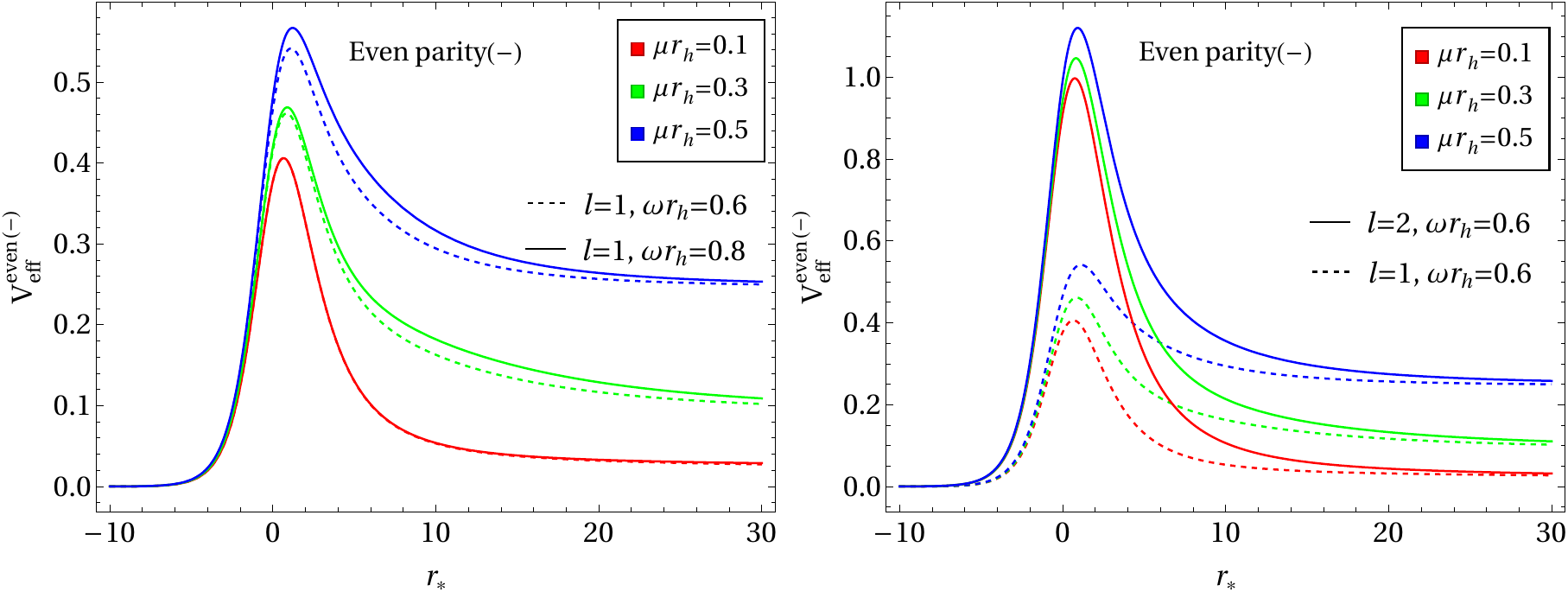}
\caption{Effective potential barrier for the negative even parity mode of the Proca field as a function of $r_*$. In the {\bf left panel}, the orbital number is fixed at $l=1$, while the frequencies $\omega r_h=0.6$ and $\omega r_h=0.8$ are shown by dotted and solid lines, respectively. In the {\bf right panel}, the frequency is held fixed, while the orbital numbers $l=1$ and $l=2$ are represented by dotted and solid lines, respectively. Different colours correspond to different values of the field mass $\mu$.}\label{fig:evenminus_pot}
\end{figure*}
Having $l\neq 0$ (implies $L\neq 0$), the separation constant naturally gives rise to two distinct branches. It is instructive to observe that in the massless limit, $\mu\to 0$, the negative root disappears and the corresponding mode acts as a pure gauge mode (similar to the monopole case), but represents a physical mode for $\mu\neq 0$ \cite{Hancock:2025ois}. Hence, we can associate this root with the scalar-type polarization. In the massless limit, the $(+)$ branch alone describes the even parity sector and is therefore identified as the vector-type polarization \cite{Fernandes:2021qvr}. 
 
In moving forward, we find it convenient to express the governing equation for all the even and odd parity modes in a coherent manner in Schr$\ddot{\rm o}$dinger-like form, as obtained in \eqref{eq. Monopole mode} and \eqref{eq: Odd_Parity_eom2}. To do the same for the even parity with $l\neq 0$, we rescale the radial function as
\be
\bar{R}(r) =\frac{\sqrt{q^\pm_r}}{r} R(r),
\label{eq: rescaling}
\ee
where $q^\pm_r=1+\nu^2_{\pm}r^2$, as mentioned before. Utilizing this scaling, the governing equation for the even parity sector reduces to
\be
\frac{d^2 R}{dr^2_*}+\left[\omega^2-V^{\rm e(\pm)}_{\rm eff}\right]R = 0,
\ee
with
\be
V^{\rm e(\pm)}_{\rm eff}=V^{\rm o}_{\rm eff}(r)+\pr_{r_*}\left(\frac{f}{rq^{(\pm)}_r}\right)+\left(\frac{f}{rq^{(\pm)}_r}\right)^2+\frac{2f\omega \nu_\pm}{q^{(\pm)}_r}.
\ee
As it turns out, in contrast to the monopole and odd parity case, the even parity effective potential for non-zero $l$ explicitly depends on the frequency $\omega$. Although for a rotating BH this is an unusual occurrence \cite{Benone_2019, Leite:2017zyb}, in the present static BH scenario, it similarly originates from the decoupling of the equation of motion. However, the definition captures all the parameter dependence to study the transmission properties in the even parity sector.

In Fig.~\ref{fig:evenplus_pot} and \ref{fig:evenminus_pot}, we illustrate the behaviour of the effective potential for the positive even parity (vector-type, $e(+)$) mode and the negative even parity (scalar-type, $e(-)$) mode, respectively. As is evident from these figures, the height of the potential barrier generally increases with increasing orbital number $l$, similar to what is observed for the other polarizations. Concerning the frequency dependence, we see that the height of the barrier for vector-type modes decreases slightly. By contrast, for scalar-type modes, we find that the peak of the barrier initially increases a bit with the frequency. However, this increase becomes progressively weaker as we increase the frequency and eventually saturates at a high frequency. This feature is particularly evident for the lower masses, namely $\mu r_h = 0.1$ and $\mu r_h = 0.3$, for which the frequency range considered in the plots extends to larger values. This suggests that the transmission for the scalar-type mode in the even parity becomes significant only at somewhat higher frequencies than for its vector-type counterpart. Although this effect is expected to be relatively small in the greybody factors of individual modes, its consequences become more pronounced in the total ACS associated with each parity sector. Regarding the mass parameter, both scalar- and vector-type modes in the even parity sector exhibit behaviour analogous to that found in the other polarizations. 

{\it Remarks on the low frequency limit:} As far as the scattering states are concerned, i.e., $\omega>\mu$, the low frequency limit effectively corresponds to the massless limit. Therefore, in this limit, the Proca theory should reduces to the EM theory. However, a redundant gauge mode remains present and must either be fixed or discarded. The ACS for the EM theory has already been studied \cite{Crispino:2007qw}, and a closed-form analytical expression in the low-frequency regime has been obtained \cite{Page:1976df}. Hence, we do not discuss the low-frequency limit for the Proca field any further. To proceed, we fix the lowest field mass to a finite value, $\mu r_h=0.1$, and study the scattering process with $\omega r_h>0.1$. It is important to note, therefore, that whenever we refer to the low-frequency regime in the following discussion, this constraint should be kept in mind. To study the effect of varying the mass, we additionally consider $\mu r_h = 0.3$ and $\mu r_h = 0.5$, restricting our attention to scattering states with $\omega > \mu$. Using these benchmark parameter values, we analyze the transmission properties in the next section.
\section{Numerical recipe for analyzing the reflection and transmission: evaluation of greybody factor}\label{sec:numerical_recipe}
From the previous section, it is understood that the effective potential becomes constant, $\mu^2$, in the spatial infinity limit, where the spacetime becomes flat. To study the scattering process by the BH spacetime, one needs to impose the boundary conditions such that the Proca field propagates as a purely ingoing wave near the horizon and a superposition of ingoing and outgoing components near the spatial infinity (see e.g., \cite{Crispino:2007qw, Karmakar:2024hng} for EM field). The asymptotic solutions can be expressed in the following manner:
\be\label{eq: u.bc}
{\Psi^\lambda(r)=\mathcal{N}_{kl}^{\lambda} \left\{ 
        \begin{array}{l} 
           \mathcal{I}^\lambda_{k l} e^{-ik r_*}+\mathcal{R}^\lambda_{k l}e^{ik r_*},~~~ r_*\to{\infty},\\
          ~~~~\sqrt{1-\frac{\mu^2}{\omega^2}}\mathcal{T}^\lambda_{\omega l}e^{-i \omega r_*},~~~r_*\to -\infty,
        \end{array}
        \right.}
\ee
with $\lambda\equiv \{\rm m,\,o,\,e(\pm)\}$ and $\Psi^\lambda(r)=\{u_{(2)}, u_{(4)}, R\}$. We also define $k=\sqrt{\omega^2-\mu^2}$. The overall normalization constant $\mathcal{N}_{kl}^{\lambda}$ will be fixed in the next section. Whereas, the three constant coefficients,
$\mathcal{I}^\lambda_{k l}, \mathcal{R}^\lambda_{k l}$ and $\mathcal{T}^\lambda_{\omega l}$ represent the incident, reflection, and transmission coefficients, respectively, for individual parity modes, $\lambda$. 

The conserved Wronskian associated with the differential equation for all the parity sectors can be expressed as 
\be
W[\Psi^\lambda,\Psi^{\lambda*}]= {\left\{ 
        \begin{array}{l} 
           2ik(|\mathcal{I}^\lambda_{kl}|^2-|\mathcal{R}^\lambda_{k l} |^2),~~r_*\to \infty,\\
          ~~~~~2ik|\mathcal{T}^\lambda_{\omega l}|^2,~~~~~~~~~~~~~r_*\to -\infty.\\
        \end{array}
        \right.}
\ee
Importantly, the transmission coefficients, mentioned above, have been defined in such a way that the conserved Wronskian ensures (see \cite{Benone:2014qaa} for similar definition with massive scalar field)
\be
(|\mathcal{I}^\lambda_{k l}|^2-|\mathcal{R}^\lambda_{k l}|^2)=|\mathcal{T}^\lambda_{\omega l} |^2.
\ee
To this end, we are now left with the determination of the above coefficients. The first step in determining them numerically is to solve the governing equation of each parity modes outlined in the previous section, with the ingoing initial condition near the horizon, $r\to r_h(1+\epsilon)$, with $\epsilon<<1$. Then the solution for individual partial modes is evaluated at $r_\infty=200 M$, which is supposed to be far away from the BH event horizon. Then by matching the numerical solution with the asymptotic form considered for $r\to \infty$ \eqref{eq: u.bc} we determine the reflection and incident coefficients. With these coefficients, the greybody factor can be defined as
\be
\Gamma^\lambda=1-\left|\frac{\mathcal{R}_{k l}}{\mathcal{I}_{k l}}\right|^2.
\ee
This physical quantity appears in Hawking flux \cite{Page:1976df}, and also importantly, in the definition of ACS, which we define in the later sections through energy flux conservation in the BH spacetime.
\begin{figure}[t]
\centering
\includegraphics[scale=0.55]{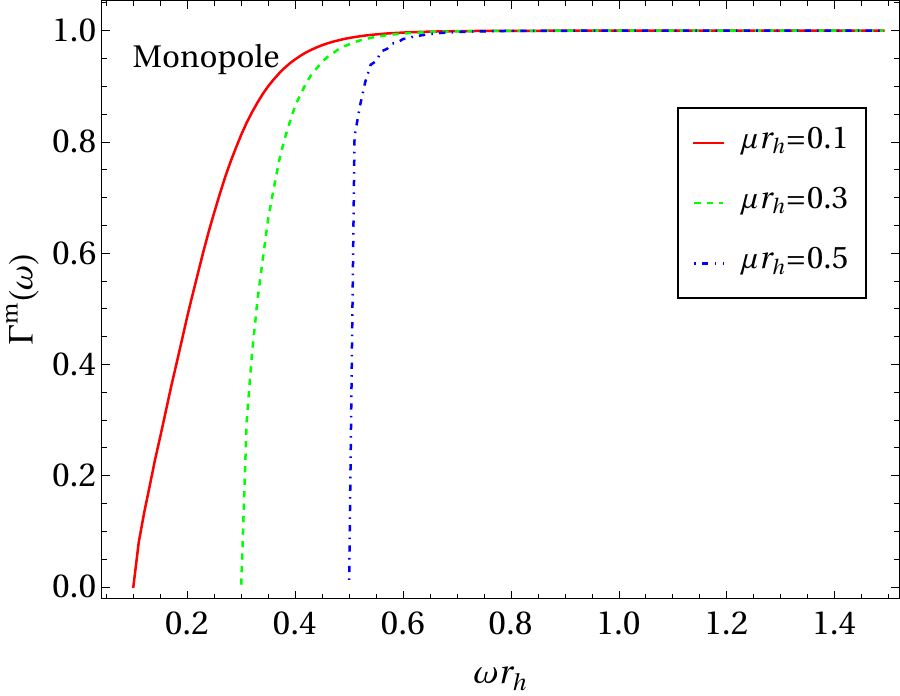}
\caption{Greybody factor for the monopole mode of the Proca field has been plotted with the frequency, $\omega$. Variations have been shown for different values of field mass $\mu$ by scaling it with the BH horizon radius, $r_h$.}\label{fig:gbody_monopole}
\end{figure}
\begin{figure*}[t] 
\centering
\includegraphics[width=\textwidth]{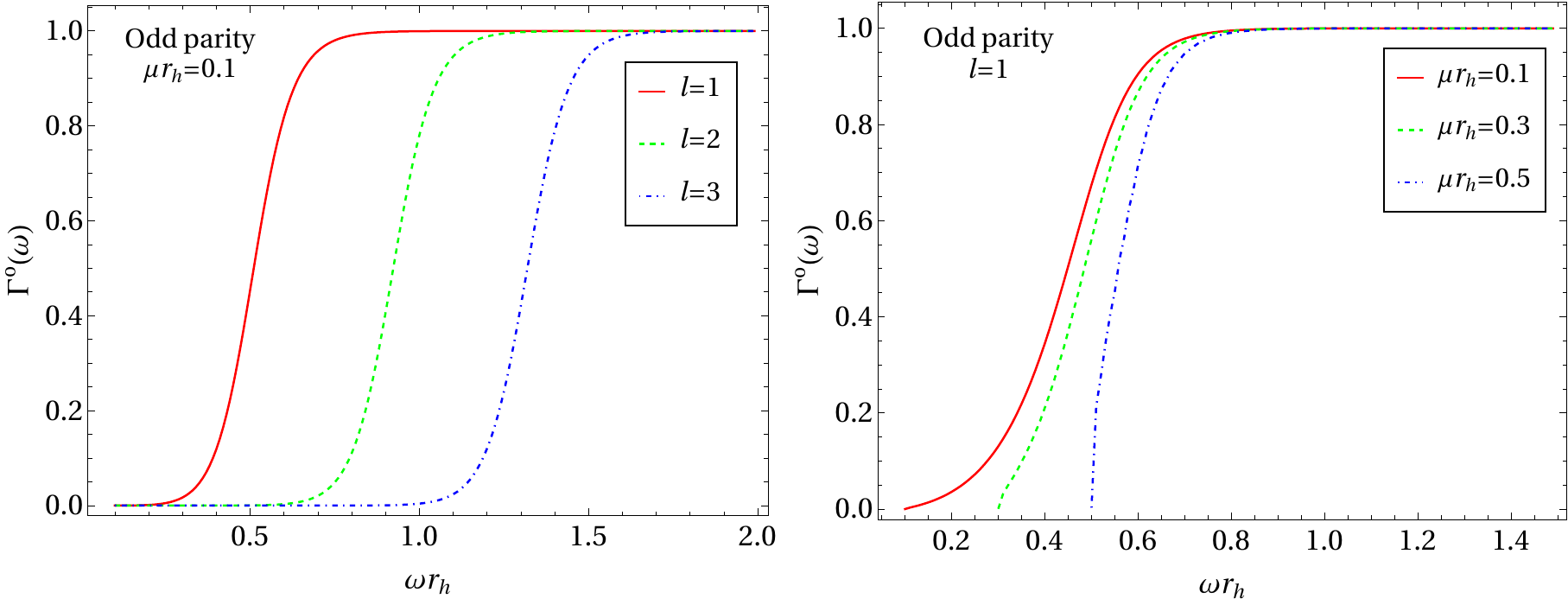}
\caption{Greybody factor for the odd parity mode of the Proca field has been plotted with the frequency,  $\omega$. In the {\bf left panel}, variations have been shown for different orbital modes, $l$, keeping the field mass fixed.  In the {\bf right panel}, variations have been shown for different values of field mass $\mu$, for a fixed orbital mode.}\label{fig:gbody_odd}
\end{figure*}
\begin{figure*}[t]
\centering
\includegraphics[width=\textwidth]{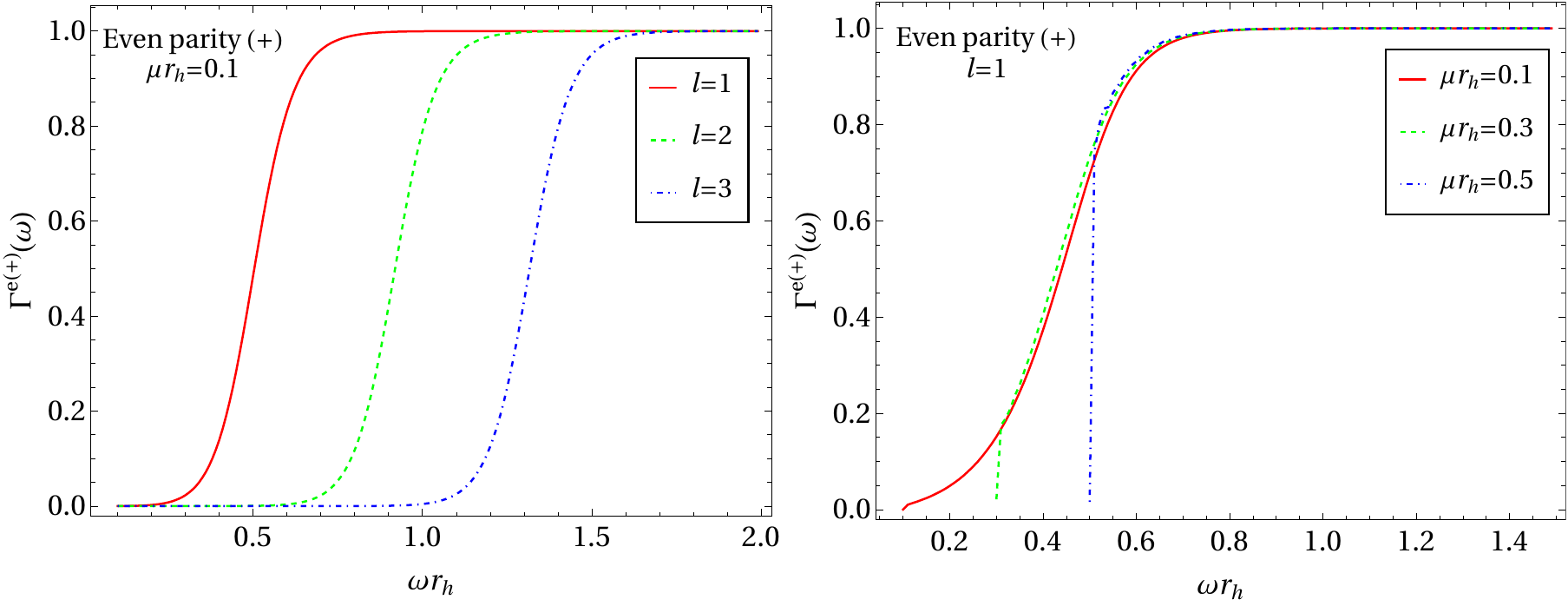}
\caption{Greybody factor for the positive even parity mode of the Proca field has been plotted with the frequency, $\omega$. In the {\bf left panel}, we have presented the results for three successive orbital modes $l$, keeping the field mass fixed. In the {\bf right panel}, keeping the orbital mode number fixed, variations have been given for different values of field mass $\mu$.}\label{fig:gbody_eplus}
\end{figure*}
\begin{figure*}[t]
\centering
\includegraphics[width=\textwidth]{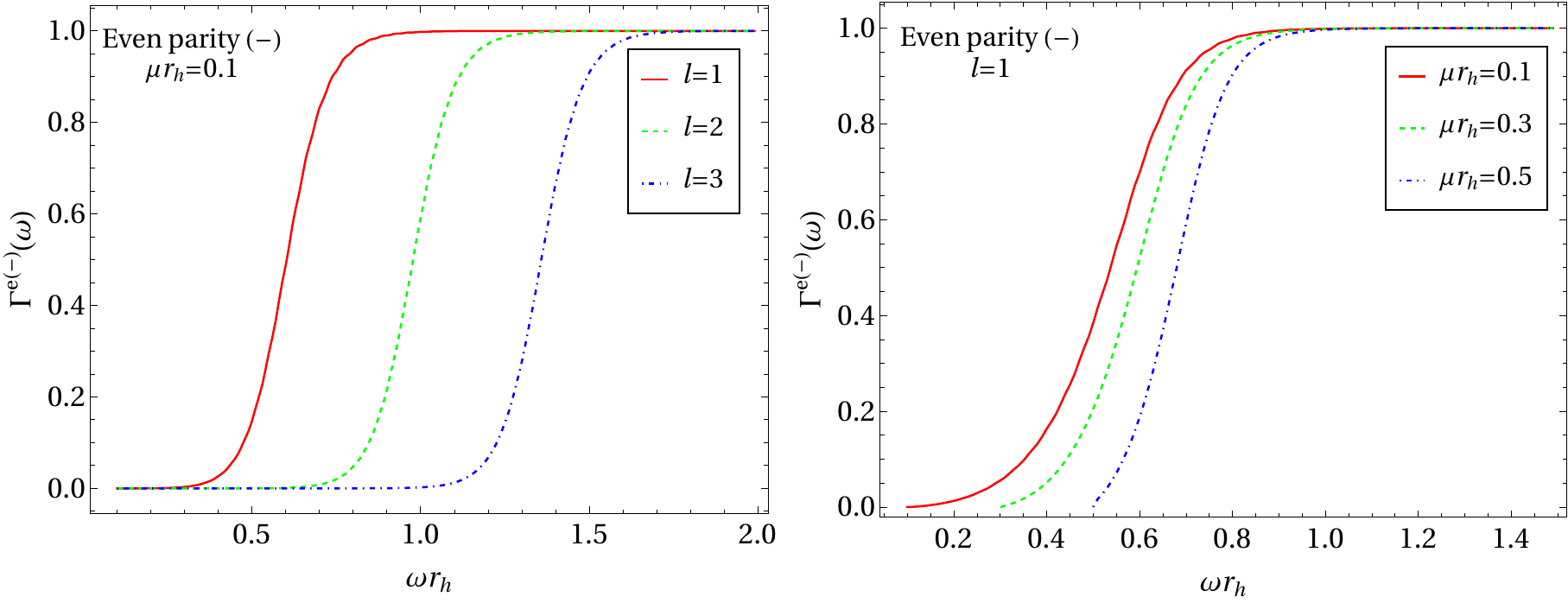}
\caption{Greybody factor for the negative even parity mode of the Proca field has been plotted with the frequency, $\omega$. In the {\bf left panel}, the behaviour has been shown for three successive orbital mode $l$, keeping the field mass fixed. In the {\bf right panel}, keeping the orbital mode number fixed, variations have been given for different values of field mass $\mu$.}\label{fig:gbody_eminus}
\end{figure*}

To understand the transmission and reflection properties, we have numerically evaluated the greybody factor for each parity sector using different values of the field mass and orbital number over a wide range of frequencies. We have presented the results in Fig.~\ref{fig:gbody_monopole}, \ref{fig:gbody_odd}, \ref{fig:gbody_eplus} and \ref{fig:gbody_eminus}, including the monopole mode ($l=0$) as well as the odd and even parity, $l\neq 0$, sector. In all of these figures, one can see that increasing the mass naturally shifts the starting frequency corresponding to non-zero transmission to higher values, as expected for the scattering states of massive fields. Another way to understand this feature is that the height of the effective potential barrier generally increases with increasing mass for all parity modes, as discussed in the previous section, thereby requiring incident waves with a larger frequency to overcome it. It is worth mentioning that similar characteristics were previously found with a massive charge scalar field in a charged BH background \cite{Benone:2014qaa}, and also present in the rotating scenario \cite{Benone_2019}. However, the magnitude of the transmission naturally depends on how much the barrier height increases. Therefore, for different masses, along with the shift in the starting frequency, the overall transmission exhibits slight mass-dependent variations that differ among the parity modes. Particularly, we observe a prominent rightward shift of the initial slope of the greybody factor for the even parity scalar-type $e(-)$ mode among other polarizations as has been shown in the right panel of Fig.~\ref{fig:gbody_eminus}. As discussed in the previous section, the initial increase in the height of the effective potential barrier for this scalar-type even parity mode also contributes to this behaviour.

The dependence of the greybody factor on the orbital mode can also be understood from the structure of the effective potential discussed in the previous section. The left panel of Fig.~\ref{fig:gbody_odd}, \ref{fig:gbody_eplus}, and Fig.~\ref{fig:gbody_eminus} illustrate the dependence of the greybody factor on the orbital number $l$. At fixed field mass, $\mu$, the peak of the potential barrier increases with $l$, and consequently, the transmission is suppressed. Accordingly, we observe that all the curves of the greybody factor systematically shift toward higher frequencies with increasing $l$, proving the dominance of the centrifugal barrier in the effective potential. 
\section{Normalization of the Proca field}\label{sec:normalization}
The usual scattering problem deals with the scenario that an incoming plane wave upon incidence, how much is reflected and how much is transmitted due to the presence of a scatterer, which, in the present case, is a spherically symmetric static BH spacetime \cite{unruh1976absorption}. For an incoming plane Proca wave at spatial infinity, the components of the field can be expressed as \cite{greiner2013field},
\be\label{eq:planeproca}
A_\alpha(t,{\bf x})\equiv \begin{pmatrix}
 \frac{k}{\mu}\\
 1\\
 i\\
 \frac{\omega}{\mu}
\end{pmatrix}
e^{-i(\omega t+kz)},
\ee
where, notably, the overall normalization does not matter, as that contribution would appear both in the numerator and denominator in the definition of ACS, as we will see later.
Recall that the scattered Proca wave in the BH spacetime happens to be spherical. To normalize such a spherical wave in a manner that its ingoing part at spatial infinity corresponds to a plane wave propagating along the z-direction, one needs to express the above Cartesian decomposition into spherical coordinates. The transformation of the Proca wave from Cartesian coordinates $(t,x,y,z)$, to spherical coordinates $(t,r,\theta,\phi)$, is done utilizing \cite{padmanabhan2010gravitation}, 
\be
A'_{\mu}(x')=\frac{\partial x^{\nu}}{\partial x'^{\mu}}A_{\nu}(x),
\ee
thereby the components in the spherical coordinates turn out to be 
\be\label{pln.sph1}
\bea
&A'_t(t,\textbf{r})=A_t(t,\textbf{x}),\\
&A'_r(t,\textbf{r})=\sin\theta e^{i\phi}A_x(t,{\xb})+\cos\theta A_z (t,{\xb}),\\
&A'_\theta(t,\textbf{r})=r\cos\theta e^{i\phi}A_x(t,{\xb})-r\sin\theta A_z (t,{\xb}), \\
&A'_\phi(t,\textbf{r})=ir\sin\theta e^{i\phi}A_x(t,{\xb}).
\eea
\ee
Where the term $A_x(t, \xb)$ appearing on the right-hand side of the above decomposition must be rewritten in spherical coordinates. To do this, we utilize the Rayleigh expansion of the spatial component of a plane wave propagating along the $z$-direction \cite{griffiths2018introduction},
\be
e^{-ik z}=e^{-i k r\cos\theta}=\sum_{l=0}(2l+1)i^l j_{l}(k r)P^0_l(\cos\theta),
\ee
where $j_l(k r)$ is the spherical Bessel function and $P^0_l(\cos\theta)$ denotes the associated Legendre function \cite{abramowitz1964handbook}.
Taking the derivative of the above equation with respect to $\theta$, the decomposition can be equivalently expressed as:
\be
e^{-ik z}=e^{-ik r\cos\theta}=\sum_{l=0}(2l+1)i^l \frac{j_{l}(k r)}{ik r}\frac{\pr_\theta P^0_l(\cos\theta)}{\sin\theta}.
\ee
With this setup in hand, the temporal and radial components of the plane proca waves in spherical coordinates turn out as (see \cite{Crispino:2007qw} for similar analysis in the case of EM  wave),
\be\label{eq: temporal_comp_decomp}
\bea
A'_t(t,\textbf{r})&=\sum_{lm} A'^{lm}_t(t,r) Y_{lm}(\theta,\phi),\\
&=i\sum_{lm}(-1)^{l}\delta_{m0}\sqrt{4\pi(2l+1)}\frac{e^{-i(\omega t+kr)}}{2\mu r}Y_{lm}\\
&~~~~~~~~~~~~+{\rm outgoing~part},\\
\eea
\ee
and
\be\label{eq: radial_comp_decomp}
\bea
&A'_r(t,\textbf{r}),\\
&=\sum_{lm} A'^{lm}_r(t,r)Y_{lm}(\theta, \phi)\\
&=\sum_{lm}\Bigg\{\delta_{m1}\sqrt{4\pi(2l+1)L}\frac{(-1)^le^{-i(\omega t+kr)}}{2k^2r^2}Y_{lm}\\
&-\delta_{m0}\sqrt{4\pi(2l+1)}\frac{\omega (-1)^{l+1}
e^{-i(\omega t+kr)}}{\mu k}\left(\frac{i}{2r}+\frac{1}{2kr^2}\right)Y_{lm}\Bigg\}\\
&~~~~~~~~~~~~+{\rm outgoing~part},
\eea
\ee
where one needs to utilize the following relation for associated Legendre's functions, $P^1_l(\cos\theta)=-\pr_\theta P^0_l(\cos\theta)$ \cite{abramowitz1964handbook}. In deriving the above expansions, we also used the following relations \cite{griffiths2018introduction}: 
\be
\bea
&j_{l}(k r)+in_{l}(k r)\sim\frac{(-\mi)^{l+1}e^{ikr}}{kr},\\
&j_{l}(kr)-i n_{l}(kr)\sim\frac{i^{l+1} e^{-ikr}}{kr},
\eea
\ee
for $r\to\infty$ or $k r>>1$, which implies 
\be
j_l(k r)\sim i^{l+1}\frac{e^{-ikr}}{2kr}+(-i)^{l+1}\frac{e^{ikr}}{2kr}.
\ee
With the above decomposition in hand, before figuring out the normalization of the Proca field solution for the incident plane wave, let us look at how the constraint condition \eqref{eq: gauge.cond} helps us in simplifying the procedure further. For the monopole mode at spatial infinity, the constraint condition reads
\be\label{eq: mono_constraint}
-\pr_tA'_t+\frac{1}{r^2}\pr_r\left(r^2A'_r\right)=0,
\ee
which is also checked to be satisfied with the spherical decomposition of the incident plane wave \eqref{eq: temporal_comp_decomp} and \eqref{eq: radial_comp_decomp}. The above constraint suggests that knowing the normalized form for one among the temporal and radial components is enough for the later computation for the monopole mode. Given this strategy, we choose to find the normalization factor of the radial component for the monopole, for which the asymptotic form reads
\be
A_r=\mathcal{N}^{\rm m}_{k}\mathcal{I}^{\rm m}_{k}\frac{e^{-i(\omega t+kr)}}{r}Y_{00}(\Omega)+{\rm outgoing~part}.
\ee
By straightaway matching this asymptotic form of the Proca fields with the spherical wave decomposition of the radial component of the incident plane wave \eqref{eq: radial_comp_decomp}, the normalization factor for the monopole mode of $u_{(2)}$ reads
\be
\mathcal{N}^{\rm m}_{k}= \sqrt{\pi} \frac{\omega}{\mu k}\frac{i}{\mathcal{I}^{\rm m}_{k}}.
\ee
Where, in identifying the normalization factor, we have neglected the terms involving higher power in $kr$.

Turning to the angular components of the fields, the spherical decomposition of the incident plane wave reads 
\be
\bea
&A'_\theta(t,{\bf r})=\sum_{l=0}i(-1)^l\sqrt{4\pi(2l+1)}\frac{e^{-i(\omega t+kr)}}{2k}\\
&~~~~~~~~~~~~~~~~~~~~~~~\times Y_{l0}\Big(\cos\theta e^{i\phi}-\frac{\omega}{\mu}\sin\theta\Big)\\
&~~~~~~~~~~~~+{\rm outgoing~part},
\eea
\ee
and
\be
\bea
&A'_\phi(t,{\bf r})=\sum_{l=0}(-1)^{l+1}\sqrt{4\pi(2l+1)}\frac{e^{-(\omega t+kr)}}{2k}Y_{l0}\sin\theta e^{i \phi}\\
&~~~~~~~~~~~~+{\rm outgoing~part}.\\
\eea
\ee
Utilizing the constraint condition \eqref{eq: gauge.cond}, the two angular components can be equivalently expressed (see Appendix \ref{sec: append_anglular_comp} for details) in the following concise manner, which will be helpful in deriving  the normalization factors,
\be
\bea
&A'_{s}(t,{\rb})=\sum_{lm}(-1)^{l+1}\delta_{m1}\sqrt{\frac{4\pi(2l+1)}{L}}\times\\
&~~~~~~\times\Big[i\pr_sY_{lm}+{\epsilon_s}^{s'}\pr_{s'}Y_{lm}\Big]\frac{e^{-i(\omega t+kr)}}{2k}+{\rm outgoing~part},
\eea
\ee
where the non-zero components of the Levi-Civita tensor on the $2$-sphere are given by ${\epsilon_\theta}^\phi=1/\sin\theta$ and ${\epsilon_\phi}^\theta=-\sin\theta$. Note that, $s$ in the subscript denotes the components corresponding to $(\theta,\phi)$ as before. Whereas, the $\delta_{m1}$ factor in the above expressions arises due to having $e^{i \phi}$ in the $A'_t$ and $A'_{\phi}$ components of the gauge field (see Eq.\ref{pln.sph1}).
To make it compatible further with the decomposition \eqref{eq: Proca_sph_decomp}, we rewrite the above expression as
\be
\bea
&A'_{s}(t,{\rb})=\frac{1}{r}\sum_{lm}(-1)^{l+1}\delta_{m1}\sqrt{4\pi(2l+1)L}\times\\
&~~~~~~~~~\times \frac{e^{-i(\omega t+kr)}}{2k}\left[ic_{(3)}Z^{(3)lm}_s+c_{(4)}Z^{(4)lm}_s\right]\\
&+{\rm outgoing~part}.
\eea
\ee
To determine the normalization factor, it is therefore imperative again to consider the asymptotic form of the angular components of \eqref{eq: Proca_sph_decomp},
\be \label{eq: Angular_comp}
\bea
&A'_{s}(t,{\rb})=\frac{1}{r}\sum_{lm}e^{-i(\omega t+kr)}\Big[\sum_{\rm e\pm}c_{(3)}L\nu_{\pm}\mathcal{N}^{e(\pm)}_{kl}\mathcal{I}^{e(\pm)}_{kl}Z^{(3)lm}_s\\
&~~~~~~~~~~~~~~~~~+c_{(4)}\mathcal{N}^{\rm o}_{kl}\mathcal{I}^{\rm o}_{kl}Z^{(4)lm}_s\Big]+{\rm outgoing~part}.\\
\eea
\ee
By comparing the preceding two asymptotic forms, we find, expectedly, that the resultant field components for Proca even parity modes are a linear combination of the even parity $(+)$ and $(-)$ branches, which can be extracted and expressed as
\be\label{eq: norm_rel_1}
\sum_{\tilde{\lambda}}L\nu_{\tilde{\lambda}}\mathcal{N}_{kl}^{\tilde{\lambda}}\mathcal{I}_{kl}^{\tilde{\lambda}}= i (-1)^{l+1} \delta_{m1} \frac{\sqrt{4\pi(2l+1)L}}{2k},
\ee
where ${\tilde{\lambda}}$ stands for $e(+)$ and $e(-)$. Also $\nu_{\tilde{\lambda}}$ signifies $\nu_+$ and $\nu_-$ for the respective branches. Nevertheless, the above single relation cannot determine the normalization factor for the individual branches, $e(+)$ and $e(-)$. At this point, we utilize the temporal component (for $l\neq 0$) of the Proca field, described by $u_{(1)}$ (see \eqref{eq: Proca_sph_decomp}), which also belongs to the even parity sector. Step by step, we find first the asymptotic form of $u_{(1)}$ at spatial infinity from \eqref{eq: fkks_ansatz} by utilizing \eqref{eq: rescaling} and \eqref{eq: u.bc}.
Then by matching the asymptotic form with the temporal component of the plane Proca wave derived above \eqref{eq: temporal_comp_decomp} we obtain another relation,
\be\label{eq: norm_rel_2}
-\sum_{\tilde{\lambda}}k \mathcal{N}_{kl}^{\tilde{\lambda}}\mathcal{I}_{kl}^{\tilde{\lambda}}= i (-1)^l \delta_{m0} \frac{\sqrt{4\pi(2l+1)}}{2\mu}.
\ee
Now, simply solving the above relation and \eqref{eq: norm_rel_1}, we determine the normalization constants in the even parity sector for both the vector- and scalar-type modes ($e(+)$ and $e(-)$), and express them as
\begin{widetext}
\be
\bea
\mathcal{N}_{kl}^{e(+)} &= \frac{1}{k L (\nu_+ - \nu_-) \mathcal{I}_{kl}^{e(+)}} \left[ k \left( i (-1)^{l+1} \delta_{m1} \frac{\sqrt{4\pi(2l+1)L}}{2k} \right) + L\nu_- \left( i (-1)^l \delta_{m0} \frac{\sqrt{4\pi(2l+1)}}{2\mu} \right) \right], \\
\mathcal{N}_{kl}^{e(-)} &= \frac{1}{k L (\nu_+ - \nu_-) \mathcal{I}_{kl}^{e(-)}} \left[ -k \left( i (-1)^{l+1} \delta_{m1} \frac{\sqrt{4\pi(2l+1)L}}{2k} \right) - L\nu_+ \left( i (-1)^l \delta_{m0} \frac{\sqrt{4\pi(2l+1)}}{2\mu} \right) \right].
\eea
\ee
\end{widetext}
Whereas for the $u_{(4)}$, i.e. for the odd parity sector, the normalization constant turns out as
\be\label{norm.fact.u4}
\mathcal{N}^{\rm o}_{kl}=\mi(-1)^{l+1}\delta_{m1}\sqrt{4\pi(2l+1)L}\frac{1}{2k\mathcal{I}^{\rm o}_{kl}}.
\ee
\section{Absorption cross section for the Proca field}\label{sec:ACS_proca}
Having obtained the normalization factor and the asymptotic nature of the field solution, one can suitably define the ACS, which is the ratio of the energy absorbed by the BH horizon per unit time to the incident energy flux density \cite{unruh1976absorption}. Furthermore, the part of the absorbed energy by the horizon can be expressed in terms of the energy measured by a static observer at spatial infinity as \cite{Benone:2014qaa, Cardoso:2019dte},
\be
\pr_t\mathcal{E}=\lim_{r\to \infty}\int {r^2}\sin\theta d\theta d\phi T_{tr}.
\ee 
Where the components of the stress-energy tensor for Proca can be obtained from \cite{greiner2013field, Hancock:2025ois},
\be
\bea
T_{\alpha\beta}&= \frac{1}{2}(F_{\alpha\mu}F^{*}_{\beta\nu} + F^{*}_{\alpha\mu}F_{\beta\nu})g^{\mu\nu} -\frac{1}{4}g_{\alpha\beta}F_{\mu\nu}F^{*\mu\nu}\\
&+ \frac{1}{2}\mu^2(A_{\alpha}A^{*}_{\beta} +A^{*}_{\alpha}A_{\beta} -g_{\alpha\beta}A_{\mu}A^{*\mu}).
\eea
\ee
Explicitly, the $T_{tr}$ component reads
\be\label{eq: Ttr}
T_{tr}=\frac{1}{2}\left[F_{ts}F^{*}_{rs'}g^{ss'}+\mu^2A_tA^{*}_r+c.c\right].
\ee
Inserting into the previous equation and integrating out the spherical part utilizing the orthonormalization condition for the spherical harmonics \eqref{eq: orthogonality_cond}, the energy flux per unit time for the monopole mode, $l=0$, appears to be
\be\label{eq: monopole_energy_flux}
\bea
\pr_t\mathcal{E}^{\rm m}&=\lim_{r\to \infty}\frac{1}{2}\sum_{lm}\Big[\mu^2u_{(1)}(t,r)u^*_{(2)}(t,r)+c.c\Big].
\eea
\ee
where $c.c$ stands for complex conjugation. As discussed in the previous section, the constraint condition \eqref{eq: mono_constraint} for monopole mode suggests that
\be
u_{(1)}(r) =\frac{i}{\omega r}\frac{d}{dr}[r u_{(2)}(r)] = \frac{i}{\omega r} [u_{(2)}(r) + r u'_{(2)}(r)].
\ee
Utilizing this formula in \eqref{eq: monopole_energy_flux}, the energy flux per unit time takes the following simplified form
\be
\bea
\partial_t\mathcal{E}^{\rm m}&=\lim_{r\to \infty}\frac{\mu^2}{\omega} {\rm Im} \left(u_{(2)}\partial_{r}u^*_{(2)}\right)\\
&=\frac{\mu^2 k}{\omega} \left|\mathcal{N}^{\rm m}_{k}\right|^2\left(\left|\mathcal{I}^{\rm m}_{k}\right|^2- \left|\mathcal{R}^{\rm m}_{k}\right|^2\right).
\eea
\ee
Before we evaluate the above expression, let us derive the form of the above quantity for the non-zero orbital mode, $l\neq 0$. This can be straightforwardly done in terms of the components by substituting the asymptotic form of the Proca fields at spatial infinity \eqref{eq: u.bc}, in the radiation flux component of the stress-energy tensor given in \eqref{eq: Ttr}. For the odd parity sector, the energy flux per unit time can be expressed as
\be
\bea
\partial_{t}\mathcal{E}^{\rm o}&=\sum_{lm}\frac{\omega k \left|\mathcal{N}^{\rm o}_{kl}\right|^2}{L}\Big(\left|\mathcal{I}^{\rm o}_{kl}\right|^2 - \left|\mathcal{R}^{\rm o}_{kl}\right|^2 \Big).\\
\eea
\ee
Whereas for the even parity sector, the above quantity turns out as
\be
\bea
\partial_t\mathcal{E}
&=\lim_{r\to\infty}\frac{1}{2}\sum_{lm}\Bigg[\left(-i\omega L +\frac{2i\omega L}{q_r} -\frac{i\omega}{q_r}(L + \mu^2 r^2)\right)\\
&~~~~~\times\left(\bar R \, \partial_r \bar R^* - \bar R^* \partial_r \bar R\right)\Bigg].
\eea
\ee
Here we utilize the scaling transformation defined in \eqref{eq: rescaling} for which the differential term in the flux transforms as
\be
\bar{R} \partial_r \bar{R}^* - \bar{R}^* \partial_r \bar{R} = \left( \frac{q_r}{r^2} \right) \left( R \partial_r R^* - R^* \partial_r R \right).
\ee
Substituting this mapping into the expression for the energy flux and combining the algebraic bracket over the common denominator $q_r = 1 + \nu_\pm^2 r^2$, we obtain
\be
\partial_t\mathcal{E}^{e(\pm)} = \lim_{r\to\infty} \frac{1}{2} \sum_{lm} \Big[ -i\omega (L\nu_\pm^2 + \mu^2) \Big] \left( R \partial_r R^* - R^* \partial_r R \right).
\ee
Evaluating the limit at spatial infinity with the asymptotic incoming and outgoing physical states defined in \eqref{eq: u.bc}, one obtains the final expression for the energy flux of the even parity modes as,
\be
\bea
\partial_t\mathcal{E}^{e(\pm)}&= \sum_{lm} \omega k (L\nu_\pm^2 + \mu^2) \left|\mathcal{N}_{kl}^{e(\pm)}\right|^2\\
&~~~~\times\left( \left|\mathcal{I}^{e(\pm)}\right|^2 - \left|\mathcal{R}^{e(\pm)}\right|^2 \right).
\eea
\ee
Now, the incident energy density current for the Proca plane wave \eqref{eq:planeproca} turns out to be $3\omega k$. Hence, the total ACS can be expressed as,
\be\label{acs.def}
\sigma=\frac{\pr_t\mathcal{E}}{3\omega k}=\sum \sigma_l,
\ee
where $\sigma_l$ represents the partial ACS. 
The partial ACS for the monopole mode reads
\be\label{eq: sigma_m}
\sigma^{\rm m}=\frac{\pi}{3k^2} \left(1- \left|\frac{\mathcal{R}^{\rm m}_{k}}{\mathcal{I}^{\rm m}_{k}}\right|^2\right)=\frac{\pi}{3k^2}\Gamma^{\rm m}(\omega).
\ee
\begin{figure}[t!]
    \centering
\includegraphics[scale=0.55]{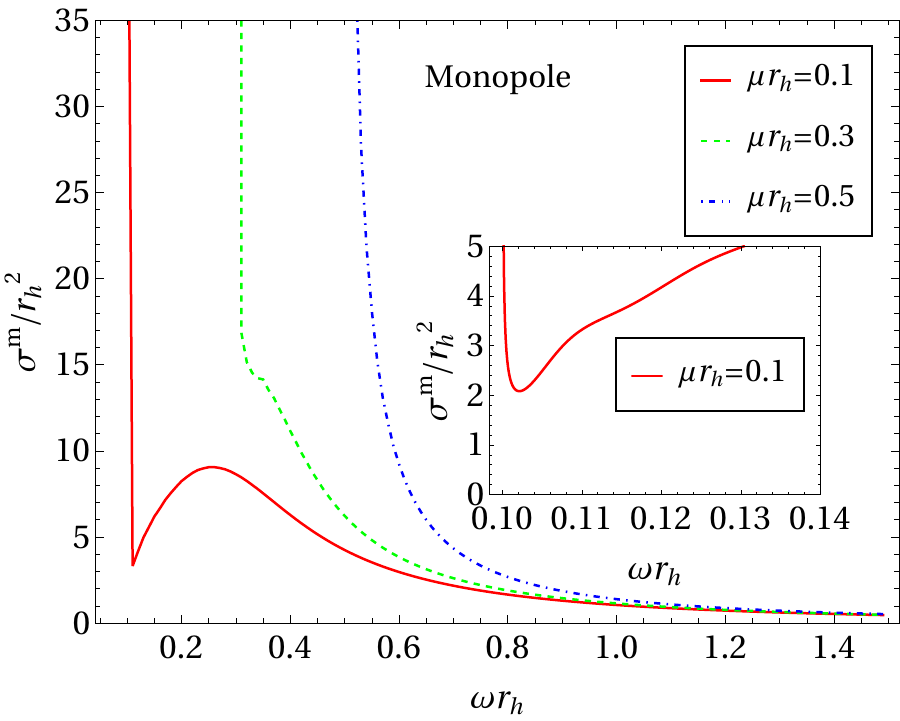}
\caption{Partial ACS frequency spectrum has been plotted considering only the monopole mode. The inset figure is the enlarged version of the monopole ACS for field mass $\mu r_h=0.1$. This shows that the ACS smoothly diverges similar to higher masses, albeit within a much smaller frequency scale, $\delta\omega \sim 0.01/r_h$.} \label{fig: monopole_acs}
\end{figure}
\begin{figure}[t!]
    \centering
\includegraphics[scale=0.55]{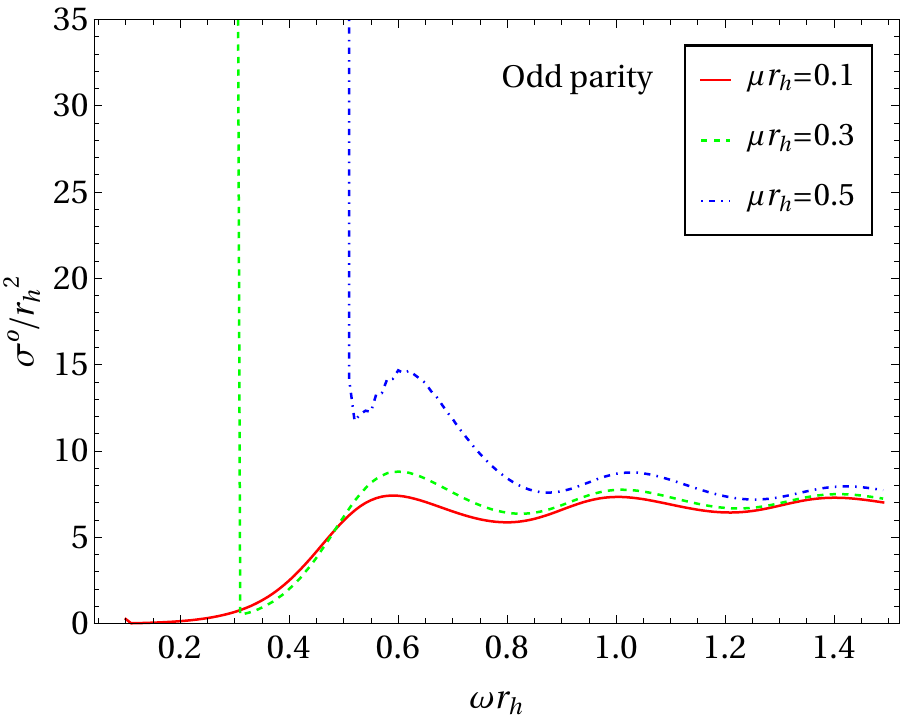}
\caption{Total ACS frequency spectrum has been plotted considering only the odd parity sector by summing over the orbital modes, up to $l=6$.}\label{fig: odd_acs}
\end{figure}
\begin{figure*}[t]
\centering
\includegraphics[width=\textwidth]{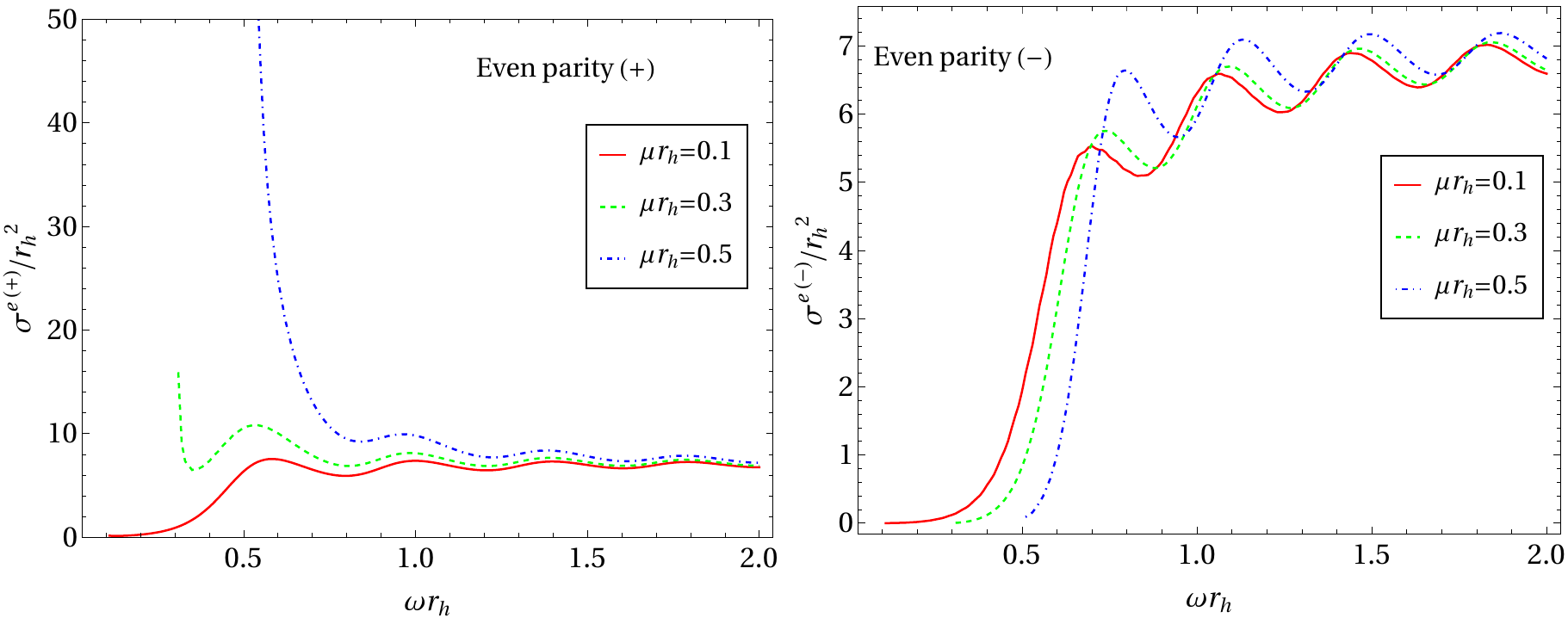}
\caption{The total ACS frequency spectrum for the even parity modes has been plotted. In the {\bf left panel}, the results for the vector-type ($e(+)$) modes have been shown. In the {\bf right panel}, the results for the vector-type ($e(-)$) modes have been shown. Variations have been presented for different values of the field mass.}\label{fig: acs_even_parity}
\end{figure*}
Whereas, the total ACS for the odd parity mode can be expressed as
\be\label{eq: odd_acs_def}
\sigma^{\rm o}=\sum_l\frac{\pi(2l+1)}{3k^2}\left(1- \left|\frac{\mathcal{R}^{\rm o}_{kl}}{\mathcal{I}^{\rm o}_{kl}}\right|^2 \right)=\sum_l\frac{\pi(2l+1)}{3k^2}\Gamma^{\rm o}(\omega).
\ee
Similarly, for the even parity modes, the total ACS for $e(\pm)$ modes reads
\be
\bea
\sigma^{e(+)} &= \sum_{l=1}^{\infty} \frac{\pi(2l+1)}{3k^2} \left[ \frac{(L\nu_+^2 + \mu^2)(L\nu_-^2 + \mu^2)}{L \mu^2 (\nu_+ - \nu_-)^2} \right] \Gamma^{e(+)}, \\
    \sigma^{e(-)} &= \sum_{l=1}^{\infty} \frac{\pi(2l+1)}{3k^2} \left[ \frac{(L\nu_-^2 + \mu^2)(L\nu_+^2 + \mu^2)}{L \mu^2 (\nu_+ - \nu_-)^2} \right] \Gamma^{e(-)}.
\eea
\ee
using \eqref{eq: nuplusminus}, the terms in square brackets evaluate to unity. Therefore the total ACS for $e(\pm)$ modes simplify to
\be\label{eq:ACS even pm}
\bea
\sigma^{\rm e(\pm)}=\sum_{l}\frac{\pi(2l+1)}{3k^2}\Gamma^{\rm e(\pm)}(\omega).
\eea
\ee
The above expressions for the ACS of the different parity modes can be straightforwardly evaluated with the numerical values of the greybody factor, which we have discussed in \Cref{sec:numerical_recipe}. Concerning the summation over the orbital modes, $l$, it is important to realize, as has also been discussed in \Cref{sec:proca_eqns} and \Cref{sec:numerical_recipe}, that there will be less transmission due to the increase of the effective potential with $l$. Hence, we find it sufficient to consider up to $l=6$ in the summation, beyond which the contribution to the total cross section becomes negligible, and the result effectively converges to the classical capture cross section (see Appendix \ref{sec: append_CCS} for the details). 

We first present the ACS for individual parity modes, specifically, for the monopole mode ($l=0$) in Fig. \ref{fig: monopole_acs}, and by summing over the orbital modes, $l$, for odd and even parity (for $l\neq 0$) in Figs.~\ref{fig: odd_acs} and \ref{fig: acs_even_parity}, respectively. In Fig. \ref{fig: monopole_acs}, for the monopole mode, we have illustrated the behaviour for different values of the mass parameter, $\mu$. Restricting our attention to scattering states, the curves are shown only for $\omega>\mu$, with the starting frequency fixed at $\omega_0=\mu+10^{-5}/r_h$, for all the ACS plots. For the lowest mass $\mu r_h=0.1$, the cross-section first increases, reaching a maximum, and then decreases monotonically in the higher frequency regime. Due to the absence of a centrifugal barrier, particularly for this mode, the magnitude of the greybody factor, and hence the cross section, remains significant in the low frequency regime. This is approximately a similar feature to that found in the partial ACS ($l=0$) for the scalar field \cite{Das:1996we}. We should emphasize that this similarity is due to the presence of the monopole mode in the scalar field, for which, even though the potential is slightly different as discussed in \Cref{sec:proca_eqns}, the absence of the centrifugal barrier makes the cross section to behave qualitatively in a similar manner (see also Section II.C of \cite{Rosa:2011my} for further discussion of this similarity). 

As $\omega\to\mu$, the cross section smoothly diverges due to the $1/k^2$ factor in the definition of ACS (see Eq.~\eqref{eq: sigma_m}, \eqref{eq: odd_acs_def} and \eqref{eq:ACS even pm}). Since this behaviour is confined to the threshold, the cross section remains finite throughout the scattering regime ($\omega>\mu$). On the other hand, when the transmission becomes very less, as occurs at low frequencies, the divergence also emerges smoothly, similar to the case of higher masses but at a much smaller scale {\footnote{We thank Sam R. Dolan for bringing this point to our attention.} in the frequency range shown in the inset of Fig.~\ref{fig: monopole_acs}. To avoid numerical instability, when the transmission becomes nearly zero for frequencies away from the threshold ($\omega=\mu$), we exclude the point of divergence. This can be seen in Fig.~\ref{fig: odd_acs} and Fig.~\ref{fig: acs_even_parity}. It is important to realize that the divergence point is independent of the parity modes for non-zero transmission and depends solely on the mass and the frequency of the fields. As far as the total ACS is concerned, these points are already taken care of by the monopole mode.

In Fig.\ref{fig: odd_acs} we compare the total ACS for the odd parity sector between different values of the field mass, $\mu$. The absorption starts from small values near the threshold $\omega \simeq \mu$ and rises rapidly. For very low frequency, which is attainable by the lowest mass, $\mu r_h=0.1$, as the transmission becomes almost zero, even with $1/k^2$ factor in the definition of ACS, the spectrum effectively begins at a vanishingly small cross section. For higher masses, as the starting frequency is higher, it leads to finite transmission. As a result, the $1/k^2$ factor leads to a huge jump in the cross section. On the other hand, in the high-frequency regime, higher modes become able to be transmitted, although with smaller magnitudes. The superposition of these partial-wave contributions causes the total cross section to exhibit damped oscillations around the classical capture cross section, which will be mentioned below.

In Fig.~\ref{fig: acs_even_parity}  we present the total ACS for the even parity sector, separated into the vector-type, $e(+)$ modes (left panel) and the scalar-type $e(-)$ modes (right panel). The vector-type mode closely resembles the odd parity case, exhibiting the same $1/k^2$-driven threshold spikes, high-frequency damped oscillations, and convergence to a constant value. Whereas for the scalar type mode, $e(-)$, as the transmission is very low, one can notice the absence of the huge spikes in the low-frequency regime. With the increase of the mass, the spectrum follows a similar nature, only the starting points shift to the higher frequency as expected from our earlier discussion of the corresponding greybody factor \Cref{sec:numerical_recipe}. Importantly, the appearance of the scalar type branch in the even parity sector and its associated non-zero absorption spectrum distinguish the Proca field from the EM theory \cite{Crispino:2007qw, Karmakar:2024hng}.
\begin{figure}[t!]
    \centering
\includegraphics[scale=0.45]{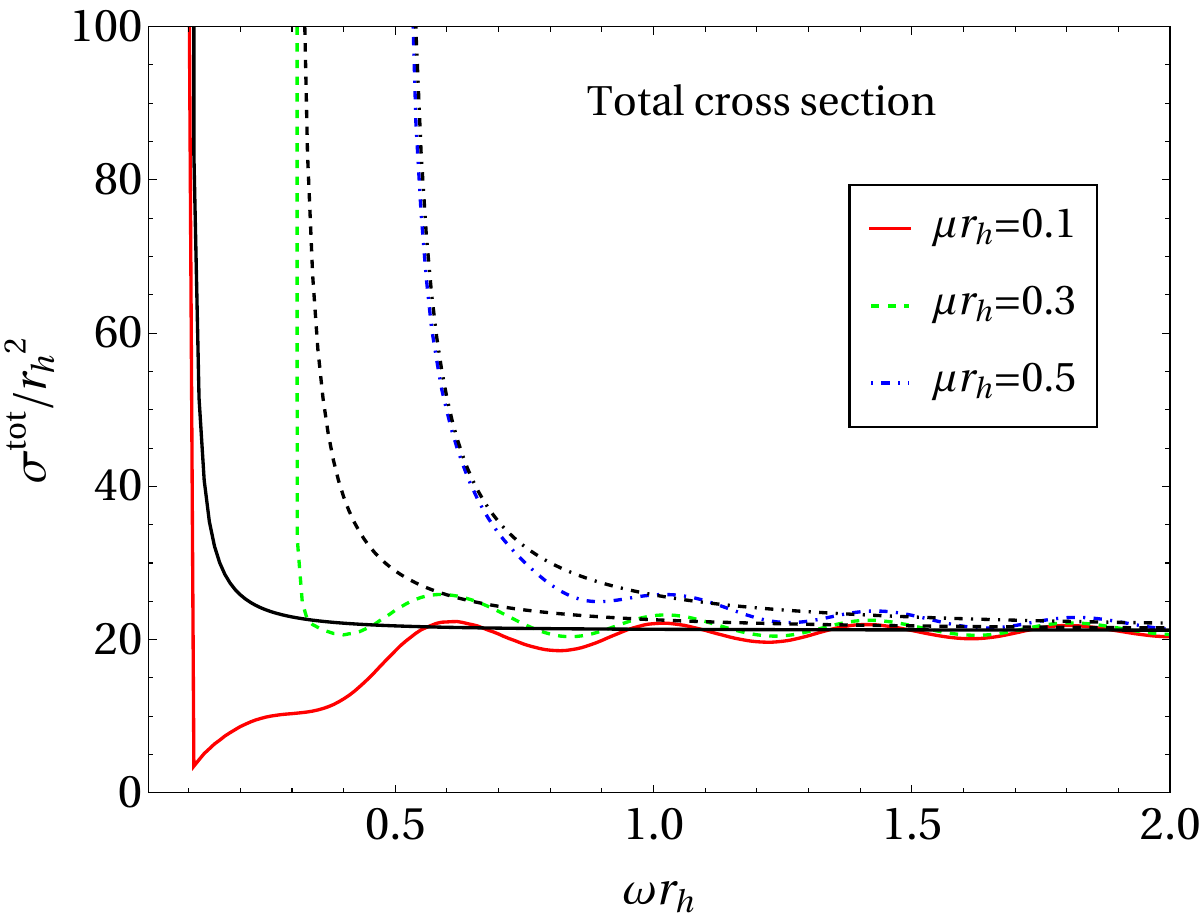}
\caption{Total ACS, combining the contribution of all the parity modes, including the monopole mode, and summing over the orbital modes, up to $l=6$,  has been plotted. The red, green, and blue lines are our numerical results, whereas the black lines are the analytical estimate of the high-frequency capture cross section.}\label{fig: tot_acs}
\end{figure}

Finally, by combining the contributions from all polarization modes, we obtain the total ACS shown in Fig.~\ref{fig: tot_acs}. The mass-dependent cross section clearly reflects the behaviour of the individual parity modes, most notably the shift of the spectrum toward higher frequencies with increasing field mass and the enhancement of the cross section in the low-frequency regime. In contrast to the EM theory, for which the total ACS vanishes in the low-frequency limit \cite{Crispino:2007qw}, the Proca-field spectra begin at finite values, apart from the divergence that occurs as $\omega\to\mu$. The smoothness of the divergence is already discussed before, and illustrated in Fig.~\ref{fig: monopole_acs}. On the other hand, in the high-frequency regime, the total ACS is expected to approach the classical capture cross section (see Appendix \ref{sec: append_CCS} for a detailed discussion). Indeed, our numerical results exhibit excellent agreement with this expectation, particularly at high frequencies, as illustrated by the black solid lines in Fig.~\ref{fig: tot_acs}.
\section{Conclusion and outlook}\label{sec:conclusion}
In the present work, we have investigated the transmission properties of the Proca field and evaluated the corresponding ACS of a Schwarzschild BH. While the absorption and transmission properties of massless fields \cite{Crispino:2009xt, Liao:2012bt}, as well as those of massive scalar \cite{Benone:2014qaa} and fermionic fields \cite{Dolan:2006vj, Doran:2005vm}, have been extensively studied in the Schwarzschild background, the corresponding scattering problem for the massive spin-1 fields has remained an open challenge difficulties associated with the separability of the field equations. Specifically, when using vector spherical harmonics to decompose the Proca field, the mass term inherently couples the radial differential equations in the even parity sector. To overcome this, we utilize the FKKS method, which decouples the Proca equation into a set of independent Schrödinger-like equations \cite{Frolov_2018, Krtous:2018bvk}. By numerically integrating the decoupled radial equations across an arbitrary range of frequencies, we determine the reflection and transmission coefficients. Importantly, we have also developed a framework for the normalization of the Proca field within this formalism. This enables us further to derive the ACS from the conserved flux and to express it in terms of the reflection and transmission coefficients.

Our numerical results reveal several distinct, polarization-dependent features in the absorption spectrum. First, we find that the monopole mode dominates the absorption spectrum in the low-frequency regime, apart from the finite divergence that occurs as the frequency approaches the field mass, $\omega \to \mu$. However, it decays monotonically at higher frequencies, reflecting its similarity with the scalar field $l=0$ mode (see Section II.C of \cite{Rosa:2011my}). On the other hand, the odd parity modes and the vector-type branch of the even parity sector exhibit qualitatively similar behaviour, particularly for larger field masses. However, the scalar-type mode associated with this even parity sector also has a non-zero contribution in the absorption spectrum. Therefore, together with the monopole mode, which also belongs to the even parity sector, the degeneracy between the odd and the even parity polarization present in the EM theory \cite{Crispino:2007qw} is lifted.  When contributions from all the polarization modes are combined, the total ACS distinctly begins at a finite value, apart from the finite divergence at the initial point. This feature contrasts with the usual EM absorption spectrum \cite{Crispino:2007qw}. Furthermore, following a regime of damped oscillations at intermediate frequencies, the summation over all orbital modes causes the total cross section to approach and eventually saturate at the classical capture cross section in the high-frequency regime. Importantly, our numerical results for this total ACS are in excellent agreement with the analytical estimate of the classical capture cross section, particularly in the high-frequency limit.  

So far, we have discussed and developed the framework for studying the absorption phenomena of Proca field in a static BH background. Therefore, an immediate extension of the present study is to consider the analogous problem in a rotating BH spacetime. Interestingly, the decoupled FKKS equations have already been formulated for Kerr spacetime \cite{Cayuso:2019ieu, Dolan_2018, Frolov:2018ezx}. Hence, the computation of the greybody factors and the ACSs should be straightforward. Such an extension is particularly important because astrophysical compact objects are generally expected to be rotating. Furthermore, recent work has demonstrated that ultralight massive vector fields can form dark matter clouds and accrete onto Schwarzschild BHs \cite{Hancock:2025ois, Karmakar:2025drp}. Integrating the exact, spin-dependent transmission spectrum into these models will be critical for determining how BH rotation alters the accretion rates and survivability of such vector dark matter halos. Finally, the exact numerical estimation of the greybody factor is fundamental to estimating the Hawking radiation spectrum \cite{Harris:2003eg}. Therefore, incorporating the transmission properties of the massive spin-1 field is highly relevant for accurately modeling the radiation flux and evaporation dynamics, particularly for primordial BHs with masses several orders of magnitude below the solar mass \cite{Cheek:2021odj, Cheek:2022mmy, Taylor:1998dk}.
\section*{Acknowledgements} 
We would like to express our gratitude to Sam R. Dolan for providing insightful comments, which helped us improve our manuscript. We thank Chandra Prakash for various discussions on the Proca field. We also appreciate Debaprasad Maity for his continuous encouragement to work in the present direction. RK acknowledges the research group of Xian-Hui Ge for various useful discussions at the Department of Physics, Shanghai University. RK would also like to thank Md Riajul Haque for various discussions on the importance of Hawking flux and, thereby, the greybody factor in the context of primordial BHs. 

\onecolumngrid
\appendix
\renewcommand{\thesection}{\Alph{section}}
\renewcommand{\theequation}{\thesection.\arabic{equation}}
\section{The angular components}\label{sec: append_anglular_comp}
In what follows, we present in detail the decomposition of the angular components of the Proca field into two independent scalar functions. The resulting formalism is closely analogous to the standard procedure employed in the analysis EM fields  \cite{Crispino:2007qw}, which we here generalize to the Proca case. Specifically, we begin by expressing the angular component of the Proca field, $A_s$, in terms of two arbitrary scalar functions $\Phi(\theta,\phi)$ and $\Psi(\theta,\phi)$ as
\be
A_s=\pr_s \Phi+\epsilon_{ss'}\pr^{s'}\Psi,
\ee
where $\epsilon_{ss'}$ is the antisymmetric Levi-Civita tensor defined on 2-sphere. By applying the covariant derivative $\nabla_s$ compatible with the metric of the 2-sphere, we can project out these scalar functions. Taking the divergence and the 2-dimensional curl of the above equation yields the following relations
\be\label{eq: div_curl}
\epsilon^{ss'}\nabla_sA_{s'}=-\nabla^2\Psi, ~~~\nabla^s A_s=\nabla^2\Phi.
\ee
Where $\nabla^2\equiv \nabla^s\nabla_s$ represents the angular Laplacian operator. To determine the scalar function $\Phi$, we invoke the Lorenz constraint condition governing the Proca field, which can be expressed in terms of the temporal and radial components as
\be
-\pr_t A'_t+\frac{1}{r^2}\pr_r(r^2A'_r)+\frac{1}{r^2}\nabla^sA_s=0.
\ee
This constraint dictates that the divergence of the angular sector, $\nabla^s A_s = \nabla^2 \Phi$, is entirely fixed by the dynamics of the temporal and radial field components, $A'_t$ and $A'_r$. As discussed in the main text, the asymptotic forms of these components are given by
\be
\bea
&A'_t(t,\textbf{r})=i\sum_{lm}(-1)^{l}\delta_{m0}\sqrt{4\pi(2l+1)}\frac{e^{-i(\omega t+kr)}}{2\mu r}Y_{lm}+{\rm outgoing~part},\\
&A'_r(t,\textbf{r})=\sum_{lm}\Bigg\{\delta_{m1}\sqrt{4\pi(2l+1)L}\frac{(-1)^le^{-i(\omega t+kr)}}{2k^2r^2}Y_{lm}-\delta_{m0}\sqrt{4\pi(2l+1)}\frac{\omega (-1)^{l+1}
e^{-i(\omega t+kr)}}{\mu k}\left(\frac{i}{2r}+\frac{1}{2kr^2}\right)Y_{lm}\Bigg\}\\
&~~~~~~~~~~~~~+{\rm outgoing~part}.
\eea
\ee
Substituting these into the constraint condition \eqref{eq: gauge.cond}, we obtain
\be
\bea
\nabla^2\Phi&=\pr_t A'_t-\frac{1}{r^2}\pr_r(r^2A'_r)\\
&=\omega\sum_{lm}(-1)^{l}\delta_{m0}\sqrt{4\pi(2l+1)}\frac{e^{-i(\omega t+kr)}}{2\mu r}Y_{lm}\\
&-\sum_{lm}\Bigg\{-i\delta_{m1}\sqrt{4\pi(2l+1)L}\frac{(-1)^le^{-i(\omega t+kr)}}{2kr^2}Y_{lm}-\delta_{m0}\sqrt{4\pi(2l+1)}\frac{\omega (-1)^{l+1}
e^{-i(\omega t+kr)}}{2\mu r}Y_{lm}\Bigg\}\\
&=\sum_{lm}i\delta_{m1}\sqrt{4\pi(2l+1)L}\frac{(-1)^le^{-i(\omega t+kr)}}{2kr^2}Y_{lm}.
\eea
\ee
Now, utilizing the fact that $\nabla^2Y_{lm}=-LY_{lm}$, we obtain
\be
\Phi=\sum_{lm}i\delta_{m1}\sqrt{\frac{4\pi(2l+1)}{L}}\frac{(-1)^le^{-i(\omega t+kr)}}{2kr^2}Y_{lm}.
\ee
On the other hand, to find out the expression $\Psi$, we utilize the angular components directly, which, as derived in the main text, are
\be
\bea
&A'_\theta(t,{\bf r})=\sum_{l=0}i(-1)^l\sqrt{4\pi(2l+1)}\frac{e^{-i(\omega t+kr)}}{2k}Y_{l0}\Big(\cos\theta e^{i\phi}-\frac{\omega}{\mu}\sin\theta\Big)+{\rm outgoing~part},\\
&A'_\phi(t,{\bf r})=\sum_{l=0}(-1)^{l+1}\sqrt{4\pi(2l+1)}\frac{e^{-i(\omega t+kr)}}{2k}Y_{l0}\sin\theta e^{i \phi}+{\rm outgoing~part},\\
\eea
\ee
therefore, applying the curl operator as defined in \eqref{eq: div_curl}, one obtains
\be
\bea
\nabla^2\Psi&=-\epsilon^{ss'}\nabla_sA_{s'}\\
&=-\frac{1}{\sin\theta} \left(\pr_{\theta}A_{\phi} - \pr_{\phi}A_{\theta} \right)\\
&=\sum_{l=0}(-1)^{l+1}\delta_{m1}\sqrt{4\pi(2l+1)}\frac{e^{-i(\omega t+kr)}}{2k}Y_{lm}+{\rm outgoing~part}.
\eea
\ee
Again utilizing the fact that $\nabla^2Y_{lm}=-LY_{lm}$, we obtain
\be
\Psi=\sum_{l=0}(-1)^{l}\delta_{m1}\sqrt{\frac{4\pi(2l+1)}{L}}\frac{e^{-i(\omega t+kr)}}{2k}Y_{lm}+{\rm outgoing~part}.
\ee

\section{High frequency capture cross section for a massive particle}\label{sec: append_CCS}
In this appendix, we review the derivation of the geometric capture cross section for a massive particle, which determines the high-frequency limit of the ACS. The Lagrangian describing the motion of a massive test particle in the Schwarzschild BH background is usually expressed as \cite{Chandrasekhar:1985kt}
\be
2\mathscr{L}=-f(r)\dot{t}^2+f(r)^{-1}\dot{r}^2+r^2\dot{\theta}^2+r^2\sin^2\theta\dot{\phi}^2
\ee
where the overdot denotes derivative with respect to an affine parameter, which can also be considered as the proper time, having considered the massive particle. Restricting the motion to be confined in the equatorial plane, the components of the generalized momenta read
\be
\bea
&p_t=-f(r)\dot{t}=-\mathcal{E},\\
&p_r=f(r)^{-1}\dot{r},\\
&p_\phi=r^2\dot{\phi}=\mathcal{L},\\
\eea
\ee
where $\mathcal{E}$ and $\mathcal{L}$ represent the conserved energy and angular momentum respectively. The time-like geodesics followed by the massive particle can then be expressed as
\be
p_\mu p^\mu=-\mu^2,
\ee
which leads to the equation for the evolution of the areal radius as
\be
\dot{r}^2=\mathcal{E}^2-f(r)\left(\mu^2+\frac{\mathcal{L}}{r^2}\right)\equiv V_r(r).
\ee
Now, for the massive particle, the impact parameter is defined as $b=\mathcal{L}/\sqrt{\mathcal{E}^2-\mu^2}$ \cite{goldstein2002classical}. Whereas, the capture cross-section of the timelike geodesic is related to the critical impact parameter, $b_c$, specifically, $\sigma_{\rm hf}=\pi b^2_c$. The critical impact parameter can be found by analyzing the unstable circular timelike orbit, which is characterized by the following conditions 
\be
V_r(r)=0, ~~~\frac{d V_r(r)}{dr}=0.
\ee
These conditions lead to 
\be\label{eq: uco_cond}
\mathcal{E}^2-f(r_c)\left(\mu^2+\frac{\mathcal{L}}{r^2_c}\right)=0,~~~f'(r_c)\left(\mu^2+\frac{\mathcal{L}^2}{r^2_c}\right)-\frac{2f(r_c)\mathcal{L}^2}{r^3_c}=0.
\ee
where $r_c$ is the radius corresponding to the unstable circular timelike orbit, which can be determined by solving the following equation, extracted from the above conditions  
\be
r^2_c\left(\frac{\mathcal{E}^2}{\mu^2}-1\right)+\left(4-\frac{3\mathcal{E}^2}{\mu^2}\right)Mr_c-4M^2=0.
\ee
The roots of this quadratic equation are given by 
\be
r_c=\frac{M}{2\left(\frac{\mathcal{E}^2}{\mu^2}-1\right)}\left[3\frac{\mathcal{E}^2}{\mu^2}-4\pm \frac{\mathcal{E}}{\mu}\sqrt{\frac{9\mathcal{E}^2}{\mu^2}-8}\right].
\label{eq:rc}
\ee
Notably, the positive root provides the physically unstable timelike orbit, given $\mathcal{E}\geq \mu$. Now the critical impact parameter can be expressed utilizing \eqref{eq: uco_cond} as
\be
b_c=\frac{1}{\sqrt{1-\mu^2/\mathcal{E}^2}}\frac{\mathcal{L}}{\mathcal{E}}\Bigg|_{r=r_c}=\frac{1}{\sqrt{1-\mu^2/\mathcal{E}^2}}\sqrt{\frac{f'(r)r^3}{2f(r)^2}}\Bigg|_{r=r_c}.
\ee
For the Schwarzschild metric, this reduces to
\be
b_c^2 = \left(\frac{\frac{\mathcal{E}^2}{\mu^2}}{\frac{\mathcal{E}^2}{\mu^2}-1}\right)\frac{M r_c^3}{(r_c - 2M)^2}.
\ee
Finally by substituting the physical root $r_c$ from \eqref{eq:rc} into the above expression, the high-frequency capture cross section is obtained as
\be\label{eq: HF ACS}
 \sigma_{\rm hf}=\frac{\pi M^2}{2} \frac{ \left( 3\mathcal{E}^2 - 4\mu^2 + \mathcal{E}\sqrt{9\mathcal{E}^2 - 8\mu^2} \right)^3 }{ \left(\mathcal{E}^2 - \mu^2\right)^2 \left( \sqrt{9\mathcal{E}^2 - 8\mu^2} - \mathcal{E} \right)^2 } .
\ee
By expressing the test particle mass in terms of its asymptotic velocity at spatial infinity, $v^2 = 1-(\mu / \mathcal{E})^2$, we can rewrite \eqref{eq: HF ACS} entirely in terms of the dimensionless velocity parameter $v$. Substituting $\mu^2 = \mathcal{E}^2 (1-v^2)$ into \eqref{eq: HF ACS} and factoring out the energy terms yeilds
\be
\sigma_{\rm hf}=\frac{\pi M^2}{2 v^4}\left[8 v^4 + 20 v^2 -1  + \left(1+ 8v^2\right)^{3/2}\right].
\ee
This expression of the classical capture cross section previously obtained for massive particles in \cite{Dolan:2006vj}.

\twocolumngrid
\bibliography{Bibliography}
\end{document}